\documentclass[acmsmall]{acmart}

\AtBeginDocument{%
  }

\usepackage{array}  
\usepackage{caption}
\usepackage{rotating}
\usepackage{makecell}
\usepackage[breakable]{tcolorbox}
\usepackage{longtable}
\usepackage{bbding}
\usepackage{multirow}
\usepackage{etoolbox} 
\usepackage{amsmath}
\usepackage{colortbl}
\usepackage{enumitem}
\usepackage{pifont}
\usepackage{float}    
\usepackage{subfig}   
\usepackage{overpic}  
\usepackage{booktabs}
\usepackage{siunitx}

\usepackage{listings}
\usepackage{xcolor}
\definecolor{codegreen}{rgb}{0,0.6,0}
\definecolor{codegray}{rgb}{0.5,0.5,0.5}
\definecolor{codepurple}{rgb}{0.58,0,0.82}
\definecolor{shallowred}{rgb}{1,0.1,0.1}
\definecolor{verylightgray}{rgb}{0.97, 0.97,0.97}

\lstdefinelanguage{Solidity}{
    keywords=[1]{assembly, assert, balance, break, call, callcode, case, catch, class, constant, continue, constructor, contract, debugger, default, delegatecall, delete, do, else, emit, event, experimental, export, external, false, finally, for, function, gas, if, implements, import, in, indexed, instanceof, interface, internal, is, length, library, log0, log1, log2, log3, log4, memory, modifier, new, payable, pragma, private, protected, public, pure, push, require, return, returns, revert, selfdestruct, send, solidity, storage, struct, suicide, super, switch, then, this, throw, transfer, true, try, typeof, using, value, view, while, with, addmod, ecrecover, keccak256, mulmod, ripemd160, sha256, sha3},
    keywordstyle=[1]\color{blue}\bfseries,
    keywords=[2]{address, bool, byte, bytes, bytes1, bytes2, bytes3, bytes4, bytes5, bytes6, bytes7, bytes8, bytes9, bytes10, bytes11, bytes12, bytes13, bytes14, bytes15, bytes16, bytes17, bytes18, bytes19, bytes20, bytes21, bytes22, bytes23, bytes24, bytes25, bytes26, bytes27, bytes28, bytes29, bytes30, bytes31, bytes32, enum, int, int8, int16, int24, int32, int40, int48, int56, int64, int72, int80, int88, int96, int104, int112, int120, int128, int136, int144, int152, int160, int168, int176, int184, int192, int200, int208, int216, int224, int232, int240, int248, int256, mapping, string, uint, uint8, uint16, uint24, uint32, uint40, uint48, uint56, uint64, uint72, uint80, uint88, uint96, uint104, uint112, uint120, uint128, uint136, uint144, uint152, uint160, uint168, uint176, uint184, uint192, uint200, uint208, uint216, uint224, uint232, uint240, uint248, uint256, var, void, ether, finney, szabo, wei, days, hours, minutes, seconds, weeks, years},	
    keywordstyle=[2]\color{codepurple}\bfseries,
    keywords=[3]{block, blockhash, coinbase, difficulty, gaslimit, number, timestamp, msg, data, gas, sender, sig, value, now, tx, gasprice, origin},	
    keywordstyle=[3]\color{violet}\bfseries,
    basicstyle=\fontfamily{zi4}\selectfont\scriptsize,
    sensitive=true,
    comment=[l]{//},
    morecomment=[s]{/*}{*/},
    commentstyle=\color{codegreen},
    stringstyle=\color{violet},
    morestring=[b]',
    morestring=[b]"
}
\lstdefinelanguage{Python}{
    keywords={and, as, assert, async, await, break, class, continue, def, del, elif, else, except, False, finally, for, from, global, if, import, in, is, lambda, None, nonlocal, not, or, pass, raise, return, True, try, while, with, yield},
    keywordstyle=\color{blue}\bfseries,
    keywords=[2]{abs, all, any, ascii, bin, bool, bytearray, bytes, callable, chr, classmethod, compile, complex, delattr, dict, dir, divmod, enumerate, eval, exec, filter, float, format, frozenset, getattr, globals, hasattr, hash, help, hex, id, input, int, isinstance, issubclass, iter, len, list, locals, map, max, memoryview, min, next, object, oct, open, ord, pow, print, property, range, repr, reversed, round, set, setattr, slice, sorted, staticmethod, str, sum, super, tuple, type, vars, zip, __import__},
    keywordstyle=[2]\color{codepurple}\bfseries,
    keywords=[3]{__name__, __doc__, __package__, __loader__, __spec__, __annotations__, __builtins__, __file__, __cached__, __path__, __class__},
    keywordstyle=[3]\color{violet}\bfseries,
    basicstyle=\fontfamily{zi4}\selectfont\scriptsize,
    sensitive=true,
    showstringspaces=false,
    comment=[l]{\#},
    morecomment=[s]{"""}{"""},
    morecomment=[s]{'''}{'''},
    commentstyle=\color{codegreen},
    stringstyle=\color{violet},
    morestring=[b]',
    morestring=[b]",
    morestring=[b]""",
    morestring=[b]'''
}

\usepackage{cleveref}

\newcolumntype{B}{>{\color{black}}c}
\newcommand{\revised}[1]{\textcolor{black}{#1}}

\usepackage{cleveref}
\usepackage{colortbl}
\definecolor{graya}{HTML}{e6e6e6}
\def\mybar#1{
   {\color{black}\rule{\fpeval{#1/\percentscale*\barwidth} cm}{\barheight}\color{graya}\rule{\fpeval{(\percentscale-#1)/\percentscale*\barwidth} cm}{\barheight}} 
}
\newcommand{\barwidth}{0.4} 
\newcommand{\barheight}{4pt} 
\newcommand{\percentscale}{100} 
\setcopyright{rightsretained}
\acmDOI{10.1145/3660772}
\acmYear{2024}
\copyrightyear{2024}
\acmSubmissionID{fse24main-p74-p}
\acmJournal{PACMSE}
\acmVolume{1}
\acmNumber{FSE}
\acmArticle{65}
\acmMonth{7}
\received{2023-09-28}
\received[accepted]{2024-04-16}

\begin{document}
\title{Static Application Security Testing (SAST) Tools for Smart Contracts: How Far Are We?}
\author{Kaixuan Li}
\affiliation{%
 \department{Shanghai Key Laboratory of Trustworthy Computing}
  \institution{East China Normal University}
  \city{Shanghai}
  \country{China}
}
\orcid{0000-0002-3517-353X}
\email{kaixuanli@stu.ecnu.edu.cn}

\author{Yue Xue}
\affiliation{%
  \institution{MetaTrust Labs}
  \city{Singapore}
  \country{Singapore}
}
\orcid{0009-0004-2141-2044}
\email{xueyue@metatrust.io}

\author{Sen Chen}
\authornote{Corresponding author.}
\affiliation{
  \department{College of Intelligence and Computing}
  \institution{Tianjin University}
  \city{Tianjin}
  \country{China}
}
\orcid{0000-0001-9477-4100}
\email{senchen@tju.edu.cn}

\author{Han Liu}
\affiliation{%
\department{Shanghai Key Laboratory of Trustworthy Computing}
  \institution{East China Normal University}
  \city{Shanghai}
  \country{China}
}
\orcid{0009-0000-8384-7933}
\email{hanliu@stu.ecnu.edu.cn}

\author{Kairan Sun}
\affiliation{%
  \institution{Nanyang Technological University}
  \city{Singapore}
  \country{Singapore}
}
\orcid{0009-0005-2510-3684}
\email{sunk0013@e.ntu.edu.sg}

\author{Ming Hu}
\affiliation{%
  \institution{Nanyang Technological University}
  \city{Singapore}
  \country{Singapore}
}
\orcid{0000-0002-5058-4660}
\email{ecnu_hm@163.com}

\author{Haijun Wang}
\affiliation{%
  \institution{Xi'an Jiaotong University}
  \city{Xi'an}
  \country{China}
}
\orcid{0009-0001-3509-3919}
\email{haijunwang@xjtu.edu.cn}

\author{Yang Liu}
\affiliation{%
  \institution{Nanyang Technological University}
    \city{Singapore}
  \country{Singapore}
}
\orcid{0000-0001-7300-9215}
\email{yangliu@ntu.edu.sg}

\author{Yixiang Chen}
\affiliation{
\department{Shanghai Key Laboratory of Trustworthy Computing}
  \institution{East China Normal University}
  \city{Shanghai}
  \country{China}
}
\orcid{0000-0003-1235-5530}
\email{yxchen@sei.ecnu.edu.cn}



\begin{abstract}
    In recent years, the importance of smart contract security has been heightened by the increasing number of attacks against them. To address this issue, a multitude of static application security testing (SAST) tools have been proposed for detecting vulnerabilities in smart contracts. 
    However, objectively comparing these tools to determine their effectiveness remains challenging. 
    Existing studies often fall short due to the taxonomies and benchmarks only covering a coarse and potentially outdated set of vulnerability types, which leads to evaluations that are not entirely comprehensive and may display bias.
    
    In this paper, we fill this gap by proposing an up-to-date and fine-grained taxonomy that includes 45 unique vulnerability types for smart contracts. Taking it as a baseline, we develop an extensive benchmark that covers 40 distinct types and includes a diverse range of code characteristics, vulnerability patterns, and application scenarios. Based on them, we evaluated 8 SAST tools using this benchmark, which comprises 788 smart contract files and 10,394 vulnerabilities. 
    Our results reveal that the existing SAST tools fail to detect around 50\% of vulnerabilities in our benchmark and suffer from high false positives, with precision not surpassing 10\%. We also discover that by combining the results of multiple tools, the false negative rate can be reduced effectively, at the expense of flagging \revised{36.77 percentage points more} functions. 
    Nevertheless, many vulnerabilities, especially those beyond Access Control and Reentrancy vulnerabilities, remain undetected. 
    We finally highlight the valuable insights from our study, hoping to provide guidance on tool development, enhancement, evaluation, and selection for developers, researchers, and practitioners. 
\end{abstract}

\maketitle
\section{Introduction}\label{intro}
As smart contracts gain prominence in the blockchain technology landscape, their secure and efficient implementation becomes increasingly crucial. These self-executing digital agreements automate transactions and enforce contract terms, but they are not immune to vulnerabilities and errors. It was reported that security vulnerabilities in smart contracts have resulted in more than \$4.75 billion in financial losses from 2012 to 2022~\cite{report2022}. Therefore, ensuring the reliability and security of smart contracts is essential to safeguard the trustworthiness of decentralized applications. 

To this end, a wide array of techniques have been proposed to detect vulnerabilities such as integer overflow/underflow. 
These approaches can be broadly classified into categories including static application security testing (SAST)~\cite{securify_pub,feist2019slither,smartcheck_pub,manticore_pub,osiris_pub,oyente_pub} performing analysis without running the programs and dynamic application security testing (DAST)~\cite{jiang2018contractfuzzer,choi2021smartian}, which involves testing the contracts with various inputs either in simulated environments before deployment or by executing the code post-deployment.
{Given the immutable nature of smart contracts, early vulnerability detection is of paramount importance~\cite{androulaki2018hyperledger}.
Compared with DAST, SAST provides more immediate and comprehensive insights, particularly during the coding phases.}

It is necessary to clearly understand the detection capability of current existing SAST tools for different stakeholders such as tool developers, researchers, and practitioners. However, it still remains challenging to objectively compare these SAST tools to determine their effectiveness. Existing studies~\cite{icse20-large,issta20-bug_injection,smartcheck_pub,feist2019slither} have demonstrated the effectiveness of some tools through a series of evaluations on their own experimental scenarios, but their limitations on evaluation settings: 
\textit{1)} \textbf{Lack of an up-to-date and fine-grained taxonomy.} 
Taxonomies~\cite{icse20-large,issta20-bug_injection,smartcheck_pub,ren_2021,icse23} are typically employed before the comparison and evaluation of SAST tools, which contributes to constructing benchmarks and testing tools. 
{Nonetheless, existing taxonomies in~\cite{dasp,icse20-large,issta20-bug_injection,jiachi-define,icse23} exhibit limitations in an outdated and coarse classification of vulnerabilities.}
For example, Durieux~\cite{icse20-large} noted that the state-of-the-art DASP Top 10~\cite{dasp} taxonomy, comprising 10 coarse vulnerability types, might not sufficiently cover all vulnerabilities affecting smart contracts. 
As for Smart Contract Weakness Classification (SWC)~\cite{swc}, despite comprising 37 entries, a considerable portion is primarily associated with code quality issues, such as \textit{SWC-103: Floating Pragma}.
{Meanwhile, DASP Top 10, proposed in 2016 and last updated in 2018, includes vulnerabilities including \textit{Short Addresses} that are now obsolete due to the introduction of corresponding security countermeasures.}
Additionally, existing taxonomies suffer from ambiguous classification and unreasonable granularity. Concretely, there are overlaps in DASP Top 10 between types, leading to ambiguities in classification (see~\S~\ref{taxonomy_construction}). 
\textit{2)} \textbf{Lack of a comprehensive benchmark.} 
The benchmarks used in existing studies~\cite{icse20-large,issta20-bug_injection} suffer from issues such as small sizes or limited vulnerability types. 
The use of these non-comprehensive benchmarks may result in varying evaluation conclusions. Ren et al.~\cite{ren_2021} concluded that evaluation results are highly reliant on benchmark suites, causing the effectiveness of Slither~\cite{feist2019slither} and  SmartCheck~\cite{smartcheck_pub} to vary across three distinct benchmarks due to the limitations in vulnerability types and code characteristics within these benchmarks. These limitations within benchmarks contribute to evaluation conclusions that may be neither objective nor fair. 
Additionally, due to the limitations within their benchmarks, while most related work~\cite{icse20-large,issta20-bug_injection} used evaluation metrics such as Precision or Recall, they generally evaluated only partial metrics, which also hinders objective evaluation conclusions.

\begin{figure}
    \centering
    \includegraphics[width=0.9\textwidth]{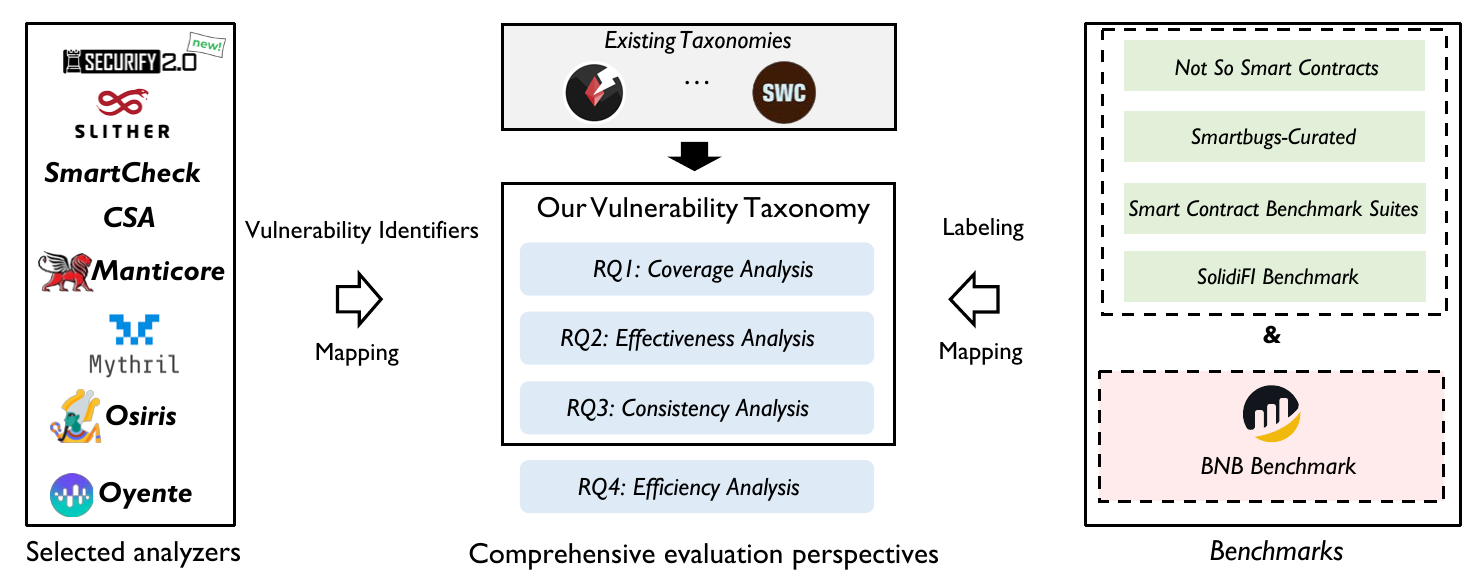}
    \caption{Overview of our study.}
    \label{fig:overview}
\end{figure}

Therefore, as shown in~\Cref{fig:overview}, to tackle the above-mentioned challenges and problems, 
we first developed {an up-to-date and fine-grained vulnerability taxonomy for smart contracts}, including 45 unique types, which is based on existing taxonomies including DASP Top 10, SWC, {the vulnerability identifiers} used in SAST tools, as well as our domain knowledge. This unified and up-to-date taxonomy facilitates the construction of our benchmark. 
We further collected and constructed a systematic and diverse benchmark suite by collecting existing high-quality datasets including~\cite{icse20-large,issta20-bug_injection,ren_2021,not-so}. Meanwhile, to enhance the benchmark's diversity and provide a more realistic representation of real-world projects, we also collected 2,941 representative industrial smart contract projects from BscScan~\cite{bscscan}. 

Notably, since some of these datasets only label partial vulnerabilities or do not disclose labeling information, we took substantial efforts (11 person-months) to manually label the ground truth by involving three security auditors to label and map them to our taxonomy at the function level. In total, our benchmark includes 788 smart contract files and 10,394 vulnerabilities (ground truth), which also covers almost all vulnerability types in our proposed taxonomy and includes a wide range of code characteristics, vulnerability patterns, and application scenarios.

Based on the new taxonomy and benchmark, we selected 8 representative SAST tools for smart contracts to compare and evaluate them. Concretely, \textit{1)} to explore and observe the detection coverage of each tool, we mapped their vulnerability identifiers to our proposed taxonomy. \textit{2)} To investigate the effectiveness of these tools, we conducted a benchmark experiment on our benchmark by adopting more fair runtime parameters and evaluation metrics, i.e., Recall, Precision, and F1-score. \textit{3)} We also performed a consistency evaluation to observe the potential for combining multiple tools. \textit{4)} Finally, we evaluated the efficiency of these tools by observing their time cost in analyzing all of the smart contracts in our benchmark. 

Through our study, we have discovered the following key findings: 
\textit{1)} CSA\footnote{The anonymous commercial static analysis tool we evaluated.} and Securify2 display higher vulnerability coverage than other tools. 
\textit{2)} Regarding effectiveness, CSA maintains its top position in both recall and precision, while Slither's performance drops due to a higher false positive rate. However, the existing tools still miss around 50\% cases in our benchmark, with precision not surpassing 10\%.  
\textit{3)} For the consistency analysis, tool combination can effectively reduce the false negatives to 29.3\%, at the expense of flagging \revised{36.77 percentage points more} functions. Additionally, vulnerabilities related to \textit{Access Control} and \textit{Reentrancy} are generally easier for tools to detect than those in the \textit{Arithmetic} category. 
\textit{4)} Our efficiency analysis shows that tools using symbolic execution take more time compared to those with static analysis techniques. Specifically, Manticore takes the longest time for analysis, and Securify2 demands more memory resources, while SmartCheck emerges as the fastest tool among those evaluated. 

In summary, our main contributions are as follows:
\begin{itemize}
    \item We proposed a new vulnerability taxonomy for smart contracts, based on existing taxonomies, the vulnerability identifiers used in SAST tools, and our domain knowledge. 
    \item  We constructed a comprehensive and diverse benchmark of $788$ smart contracts, uncovering $10,394$ vulnerabilities at the function level. The meticulous effort to establish this ground truth took \textit{11 person-months}, making it the largest benchmark dedicated to smart contract vulnerabilities to date.
    \item \revised{We conducted a benchmark experiment of 8 SAST tools using our diverse benchmark and performed a large-scale experiment on 8,981 smart contracts (i.e., $8 \times (788 + 8,981) = 78,152$ scanning tasks). This preparation and execution process took over 4 months. We performed an extensive and rigorous empirical evaluation of 8 existing SAST tools from a comprehensive perspective including coverage, effectiveness, consistency, and efficiency. 
    The study data and code are released on the website: \url{https://sites.google.com/view/sc-sast-study-fse2024/home}.} 
\end{itemize}

\section{Related Work}

\subsection{\revised{Empirical Study on SAST Tools for Smart Contracts}}
\revised{There has been extensive research conducted to evaluate security tools for smart contracts. Ghaleb and Pattabiraman~\cite{issta20-bug_injection} proposed an automated approach, SolidiFI, to evaluate static analyzers for smart contracts by injecting code defects into 50 smart contracts, introducing 9,369 security vulnerabilities, and testing the generated buggy contracts using six static analyzers. Rameder~\cite{rameder2021systematic} conducted a systematic literature and tool review, providing a comprehensive overview of tools, classifications of smart contract vulnerabilities, and detection methods including fuzzing, formal methods, and static analysis. 
Durieux et al.~\cite{icse20-large} presented an empirical evaluation of nine analyzers based on 69 annotated vulnerable smart contracts and another 47,518 contracts without ground truth. However, our work aims to conduct a systematic evaluation of static analyzers by constructing a more diverse benchmark (ground truth) and evaluating them from multi-dimensional perspectives including coverage and granularity, effectiveness, consistency, and efficiency. 
Ren et al.~\cite{ren_2021} proposed a unified standard that includes a 4-step evaluation process to eliminate bias in the assessment process. They evaluated nine representative tools based on 46,186 source-available smart contracts collected from four influential organizations. While they emphasized that experiment setups can lead to different or even incorrect conclusions during the evaluation, they did not systematically compare the detection capacities of the tools. 
Monteiro~\cite{monteiro2019study} introduced SmartBugs, an extendable execution framework that includes 47,661 Solidity smart contracts, and presented an evaluation of seven state-of-the-art tools. The tool evaluation framework they proposed contributes to our research on running security tools for Solidity smart contracts. However, the taxonomies they used are outdated or incomplete, which will be discussed in \S~\ref{taxonomy_construction}. }

\revised{
Another related work was proposed by Chen et al.~\cite{jiachi-define}, in which they collected and analyzed smart-contract-related posts and real-world smart contracts, defining 20 types of contract defects that impact smart contract quality. However, they did not compare or evaluate any security tools. We do not merely construct a systematic taxonomy for smart contract vulnerabilities and further collect a comprehensive benchmark, but evaluate the performance of eight static analyzers based on the diverse vulnerability types. 
Zhou et al.~\cite{dawnsong} constructed a DeFi reference framework that categorizes 77 academic papers, 30 audit reports, and 181 incidents, revealing differences between the academic and practitioner communities in defending and examining incidents. Their results show that DeFi security is still in its infancy and that many potential defense mechanisms require further research and implementation. 
Meanwhile, Chaliasos~\cite{defi_zhou} also focused on DeFi attacks, evaluating automated security tools and surveying 49 smart contract developers and auditors. They underlined the limited effectiveness of current tools in detecting high-impact vulnerabilities, with only 8\% of the attacks in their dataset being detected by automated tools. This emphasizes that smart contracts and DeFi security still have significant room for improvement. While reentrancy vulnerabilities can be detected, existing tools struggle to address logic-related bugs and protocol-layer vulnerabilities. 
Contrasting their work, our study focuses on vulnerabilities beyond DeFi attacks, as DeFi vulnerabilities are typically business-logic-related and considered too challenging for static analyzers to detect effectively. 
Akca et al.~\cite{esem-nonsast} evaluated various techniques for testing smart contracts including fuzzing and genetic algorithm rather than static analysis. }

\revised{Our work distinguishes itself by focusing on conducting a systematic evaluation of SAST tools for Solidity smart contracts, constructing a comprehensive benchmark, and evaluating the tools from various perspectives. We aim to provide valuable insights into the performance of these tools and contribute to the ongoing development and improvement of SAST tools in the context of smart contract security. 
In summary, our work differs from the state of the art in terms of the considered 
\textit{\textbf{(1)}} \textbf{vulnerability taxonomies} ({an up-to-date and fine-grained} vulnerability taxonomy for smart contracts), 
\textit{\textbf{(2)}} \textbf{evaluation benchmarks} (the largest benchmark covering comprehensive and diverse vulnerability types), \textit{\textbf{(3)}} \textbf{evaluation methodology} (mapping detection rules and ground truth to our taxonomy), 
\textit{\textbf{(4)}} \textbf{Finer detection code granularity} (vulnerable \textit{function-level}), 
and \textit{\textbf{(5)}} \textbf{evaluation perspectives} (coverage analysis on supported vulnerabilities, effectiveness analysis on vulnerability detection, consistency analysis among tools' focuses, and efficiency analysis). }

\subsection{\revised{SAST Tools for Smart Contracts}}
\revised{Several works focused on detecting vulnerabilities in smart contracts~\cite{liu2020towards,feist2019slither,smartcheck_pub,bose2022sailfish,securify_pub,oyente_pub,osiris_pub,manticore_pub,ghalebachecker}, with some specifically targeting certain types of vulnerabilities. Apart from the selected tools we evaluated, Liao et.al~\cite{smartdagger} proposed a framework named SmartDagger to detect cross-contract vulnerability through static analysis at the bytecode level. 
Besides, Liu et al.~\cite{liu2020towards} proposed mining past transactions of a contract to recover a probable access control model to identify potential user permission-related bugs. They implemented their role mining and security policy validation in a tool called \textsc{SPCon}. Similarly, Ghaleb et al.~\cite{ghalebachecker} introduced AChecker to detect permission-related vulnerabilities by combining data-flow analysis and symbolic execution techniques. Recently, Fang et al.~\cite{fang2023beyond} developed SoMo which leverages symbolic execution to detect \texttt{modifier} issues. }

\section{Overview}
\revised{In this section, we will introduce the overview of our study design including tool selection (\S~\ref{tool_selection}), taxonomy construction (\S~\ref{taxonomy_construction}), dataset collection (\S~\ref{datasets}), and research questions (\S~\ref{rq}), as also shown in~\Cref{fig:overview}. }

\subsection{Tool Selection}\label{tool_selection}
\revised{
To collect representative SAST tool candidates for smart contracts, we conducted a systematic literature review (SLR) as follows. \textit{1)} We first used several keywords such as ``smart contract'', ``static analysis'', and ``security tool'' to search papers published in top-tier Software Engineering, Security, and Programming Language venues in the last three years (till Sep. 2023), including FSE, ICSE, S\&P, Usenix Security, PLDI, POPL, etc. We obtained 59 research papers related to security for smart contracts as an initial paper list. 
\textit{2)} By further excluding papers that are unrelated to SAST tools, we thereby got 14 papers related. We gathered a list of 44 SAST tools. Our selection process was structured to ensure the relevance and practicality of the tools for our analysis. }

\noindent\revised{\textit{\textbf{Criterion \#1 (Availability):}} The tool should be publicly accessible. Hence, we excluded 16 tools due to unavailability (commercial or closed-source tools). Notable exclusions in this step included tools like Zeus~\cite{kalra2018zeus}, Sereum~\cite{rodler2018sereum}, and MythX~\cite{mythx}. }

\noindent\revised{\textit{\textbf{Criterion \#2 (Security related):}} We try to select tools that identify security vulnerabilities, rather than those aimed at detecting code quality issues, we thereby filtered out 2 linters including Solhint~\cite{solhint} and Ethlint~\cite{ethlint}. 
}

\noindent\revised{
\textit{\textbf{Criterion \#3 (Generalized SAST tools):}} We then narrowed our focus to ``generalized'' SAST tools since we aim to compare and evaluate tools across various vulnerability types (45 unique ones in this study). 
It required the tools to support at least \textbf{five} vulnerability types rather than specializing in a few vulnerability types, such as only aiming at few vulnerabilities within access control (e.g., AChecker~\cite{ghalebachecker}), DoS (e.g., eTainter~\cite{etainter}), state inconsistency (e.g., SailFish~\cite{bose2022sailfish}), etc. 
After this filtering, we retained 16 tools but excluded 10 tools. Note that SailFish~\cite{bose2022sailfish} was excluded since it only supports 3 vulnerability types.}

\noindent\revised{
\textit{\textbf{Criterion \#4 (CLI \& no additional input need):}} Next, due to the need for large-scale analysis, we focused on tools that support CLI and accept source code directly. Based on it, we excluded 4 tools including teEther~\cite{teEther2018}, SmarTest~\cite{so2021smartest}, NPChecker~\cite{wang2019detecting}, and SmartPulse~\cite{stephens2021smartpulse} from our list since they required additional inputs like specifications, transactions, or test cases.}

\noindent\revised{\textit{\textbf{Criterion \#5 (Popularity and relevance):}} 
In the final step, we considered several quantitative metrics that reflect their impact, adoption, and relevance in the research community and among practitioners. Specifically, we investigated \ding{172} the frequency of employed as baselines in relevant studies over the recent three years (\# Baseline), \ding{173} the citations they receive (\# Citation), and \ding{174} the number of stars in their GitHub repositories (\# Stars) till September 2023. We thereby got 7 tools that are both more popular and relevant in the field. }

\revised{
According to the unified criteria above, we finally got 7 open-source SAST tools: Securify2~\cite{securify2}, Slither~\cite{feist2019slither}, SmartCheck~\cite{smartcheck_pub}, Manticore~\cite{manticore_pub}, Mythril~\cite{mythril}, Osiris~\cite{osiris_pub}, and Oyente~\cite{oyente_pub} (see Table~\ref{tab:selected_tools}). The detailed process of each step can be accessed on our website~\cite{website-tool-selection}. 
Meanwhile, to observe the gap in effectiveness between open-source and commercial tools, we try to include commercial tools for smart contracts. However, due to the significant financial expenses, it is not feasible to run multiple commercial tools on our extensive dataset (8,981 contract files). We finally successfully obtained access to a commercial static analysis tool (CSA) through our industrial partner. }

\noindent{\textbf{Selected tools.}}
As displayed in Table~\ref{tab:selected_tools}, the selected tools leverage state-of-the-art analysis techniques and were frequently compared or evaluated in recent works~\cite{icse20-large,issta20-bug_injection,ren_2021,icse23, ghalebachecker, kalra2018zeus, bose2022sailfish, fang2023beyond} and/or are popular among practitioners\revised{~\cite{chaliasos2023smart}}. 

\begin{table}
\caption{List of selected SAST tools. \# Baseline: Tool usage as a baseline in 2021-2023. \# Stars: GitHub repository stars. \# Citation: Number of citations.}\label{tab:selected_tools}
\resizebox{0.9\columnwidth}{!}{
\begin{tabular}{@{}ccccccr@{}}
\toprule
\textbf{Technology} & \textbf{Tool} & \textbf{Analysis Level} & \textbf{\# Baseline} & \textbf{\# Citation} & \textbf{\# Stars} & \textbf{Publication} \\ \midrule
\multirow{4}{*}{\textbf{Static Analysis}} & \textbf{Securify2}  & Source Code & 7           & 649           & 529           & CCS'18\cite{securify_pub}        \\
                    & \textbf{Slither}    & Source Code & 7           & 244           & \textbf{4.5k} & WETSEB'19\cite{feist2019slither} \\
                    & \textbf{SmartCheck} & Source Code & 7           & 481           & 315           & WETSEB'18\cite{smartcheck_pub}   \\
                    & \textbf{CSA}        & Source Code & /           & /             & /             & /                                \\ \midrule
\multirow{4}{*}{\textbf{Symbolic Execution}} & \textbf{Manticore}  & Bytecode    & 7           & 185           & 3.5k          & ASE'19\cite{manticore_pub}       \\
                    & \textbf{Mythril}    & Bytecode    & \textbf{11} & /             & 3.5k          & White Paper                      \\
                    & \textbf{Osiris}     & Bytecode    & 6           & 222           & 50            & ACSAC'18\cite{osiris_pub}        \\
                    & \textbf{Oyente}     & Bytecode    & 8           & \textbf{1897} & 1.3k          & CCS'16\cite{oyente_pub}          \\ \bottomrule
\end{tabular}
}
\end{table}

\noindent\textbf{Securify2} ($\star$ 529) is a successor for Securify security scanner~\cite{securify_pub, securify}, while the latter is deprecated since 2020. In our study, we use version \texttt{v2.0}. 

\noindent\textbf{Slither} ($\star$ 4.5k) is a static analysis framework developed by Trail of Bits~\cite{trailofbits}. It converts Solidity smart contracts into an intermediate representation (IR) called SlithIR and applies Static Single Assignment (SSA)~\cite{alpern1988detecting,rwz881988wegman,cytron1989efficient} to perform data flow analysis, as well as taint tracking to extract and refine information. In our study, we use version \texttt{v0.9.3}. 

\noindent\textbf{SmartCheck} ($\star$ 315) is a static analysis tool developed by SmartDec~\cite{SmartDec}. It uses lexical and syntactical analysis performed on Solidity source code to look for vulnerabilities. We use \texttt{v2.0} of SmartCheck here.

\noindent\textbf{CSA} refers to (\underline{C}ommercial \underline{S}tatic \underline{A}nalysis), which uses static analysis technique to find vulnerabilities in smart contracts and is the commercial tool used in this study. Similar to the other tools, we make sure to use one of the latest versions of CSA. To protect the company behind this tool, we anonymize its name and neither reveal its exact version and release date nor the implemented analysis technique. 

\noindent\textbf{Manticore} ($\star$ 3.5k) is a symbolic execution tool for smart contracts. And it is integrated with Z3~\cite{z3} which is a powerful theorem prover and solver for satisfiability modulo theories (SMT)~\cite{Barrett2018} problems. We use version \texttt{v0.3.7} here.

\noindent\textbf{Mythril} ($\star$ 3.5k) is developed by ConsenSys~\cite{mythx}. It performs symbolic execution for smart contracts, relying on taint analysis, and control flow analysis of the EVM bytecode to prune the search space and to look for values that allow exploiting vulnerabilities. We use version \texttt{v0.23.15}. 

\noindent\textbf{Osiris} ($\star$ 50) was developed based on Oyente, which detects vulnerabilities in Ethereum smart contracts. Here, we use version \texttt{\#d1ecc37}. 

\noindent\textbf{Oyente} ($\star$ 1.3k) \revised{is the earliest academic tool in this domain, and it is still continuously used in evaluations of numerous academic work~\cite{icse20-large,issta20-bug_injection,ren_2021,icse23,ghalebachecker,chaliasos2023smart,xue2020cross,tse-chen,liu2018reguard,jiang2018contractfuzzer,permenev2020verx, sendner2023smarter}. 
It performs symbolic execution for vulnerability detection. And we use version \texttt{v0.2.7} here. }

\subsection{Taxonomy Construction}\label{taxonomy_construction}
Many taxonomies have been proposed~\cite{dasp,swc,defi_zhou,icse23} to facilitate vulnerability understanding, such as DASP Top 10 and SWC. However, they exhibit several limitations: 
\ding{172} {Outdated and coarse types}: Existing taxonomies may lack newly discovered vulnerability types or retain outdated types. For instance, DASP Top 10~\cite{dasp}, used by Durieux~\cite{icse20-large}, was proposed in 2016 and last updated in 2018, leading to outdated types such as the ``Short Addresses'' issue, which has been resolved by the Solidity compiler from \texttt{v0.5.0}. Moreover, SWC was established in 2017 and contains only 37 types of weaknesses, many of which focus on code quality rather than severe security risks. 
\ding{173} Ambiguous classification and unreasonable granularity: Some taxonomies exhibit overlaps between categories, leading to ambiguities in classification. For example, the taxonomy used by~\cite{dasp} shows overlaps between categories such as \textit{Access Control} and \textit{Unchecked Low-Level Calls}, which might result in confusion when categorizing vulnerabilities.  
Furthermore, the limitations in existing taxonomies make it challenging to establish a reliable benchmark for evaluating the actual performance of SAST tools or techniques in addressing smart contract vulnerabilities. 
For instance, the datasets used in~\cite{icse20-large,issta20-bug_injection}, and~\cite{monteiro2019study} are limited in benchmark sizes or vulnerability types, which lead to incomparable results and hinder the evaluation of the tools' effectiveness in vulnerability detection. 

Hence, {an up-to-date and fine-grained} taxonomy of known smart contract vulnerabilities is essential. This foundation allows for more objective and reliable comparison and evaluation of tools in the domain. 
To address this, we constructed a new vulnerability taxonomy for smart contracts from five perspectives as follows: 
\ding{172} We began by addressing {the outdated nature of existing taxonomies}, such as the DASP Top 10 and SWC. We removed outdated categories like the \textit{Short Addresses} and added newly discovered vulnerability categories \textit{Storage and Memory} to ensure a comprehensive representation of known smart contract vulnerabilities. 
\ding{173} We reorganized the taxonomy by analyzing the root causes of vulnerabilities, such as those in \textit{Block Manipulation} and \textit{Cryptographic}, to create a more structured approach in line with the ETSI smart contract standard. 
\ding{174} To resolve ambiguity and granularity issues in existing taxonomies, we combined overlapping categories and carefully analyzed their parent-child relationships. This step was to diminish ambiguity and redundancy, ensuring a clear and concise classification of vulnerabilities.
{\ding{175} Meanwhile, to ensure our taxonomy's relevance and practicality, we also considered the supported vulnerability types among state-of-the-art SAST tools by examining their {vulnerability identifiers} and aligning them with the categories in our taxonomy. This step was crucial to guarantee the applicability of our taxonomy to real-world tools and practices.} 
\ding{176} Lastly, since our primary concern is vulnerabilities in smart contracts, we removed entries related solely to code quality. 

In summary, as shown in Table~\ref{tab:taxonomy}, our taxonomy extends and refines the existing taxonomies including DASP Top 10 and SWC by addressing their limitations and reorganizing the structure for a more precise representation of smart contract vulnerabilities. 
In total, our proposed taxonomy encompasses 45 vulnerability types, organized into 7 distinct categories. Due to space limitations, details on our taxonomy can be accessed on our website~\cite{website}.

\begin{table}[]
\caption{Our proposed vulnerability taxonomy for smart contracts \revised{(\# Types indicates the number of types in each category. 
``$^*$'' indicates types not included in our benchmark. A ``\checkmark'' indicates support by tools for a given type, followed by the number of detected samples from our benchmark. \# Samples indicates the total count of vulnerability samples for each type within our benchmark.
\# SWC: the number of SWC entries mapped. \# Supported: the number of types mapped.  ).}}
\label{tab:taxonomy}
\resizebox{0.99\columnwidth}{!}{
\begin{tabular}{c|c|BBBBBBBB|B}
\hline
\textbf{Category (\# Types)} & \textbf{Type} & \makecell[b]{\rotatebox{90}{\textbf{Securify2}}} & \makecell[b]{\rotatebox{90}{\textbf{Slither}}} & \makecell[b]{\rotatebox{90}{\textbf{SmartCheck}}} & \makecell[b]{\rotatebox{90}{\textbf{CSA}}} & \makecell[b]{\rotatebox{90}{\textbf{Manticore}}} & \makecell[b]{\rotatebox{90}{\textbf{Mythril}}} & \makecell[b]{\rotatebox{90}{\textbf{Osiris}}} & \makecell[b]{\rotatebox{90}{\textbf{Oyente}}} & \makecell[b]{\rotatebox{90}{\textbf{\# Samples}}} \\ \hline
 & Arbitrary from in \texttt{transferFrom()} without msg.sender   Check &  & \checkmark (11) &  & \checkmark (15) &  &  &  &  & 17 \\
 & \cellcolor[HTML]{EFEFEF}Call to Arbitrary Addresses   with Unchecked Call data & \cellcolor[HTML]{EFEFEF} & \cellcolor[HTML]{EFEFEF}\checkmark (1) & \cellcolor[HTML]{EFEFEF} & \cellcolor[HTML]{EFEFEF}\checkmark (1) & \cellcolor[HTML]{EFEFEF} & \cellcolor[HTML]{EFEFEF} & \cellcolor[HTML]{EFEFEF} & \cellcolor[HTML]{EFEFEF} & \cellcolor[HTML]{EFEFEF}1 \\
 & Caller Not Checked & \checkmark (2) & \checkmark (5) &  & \checkmark (10) &  &  &  &  & 11 \\
 & \cellcolor[HTML]{EFEFEF}Logic Contract Could be Destructed{$^*$} & \cellcolor[HTML]{EFEFEF} & \cellcolor[HTML]{EFEFEF}\checkmark & \cellcolor[HTML]{EFEFEF} & \cellcolor[HTML]{EFEFEF}\checkmark & \cellcolor[HTML]{EFEFEF} & \cellcolor[HTML]{EFEFEF} & \cellcolor[HTML]{EFEFEF} & \cellcolor[HTML]{EFEFEF} & \cellcolor[HTML]{EFEFEF}-\\
 & Dangerous Immediate Initialization of State Variables &  & \checkmark (10) &  & \checkmark (9) &  &  &  &  & 12 \\
 & \cellcolor[HTML]{EFEFEF}Dangerous Usage of \texttt{tx.origin} & \cellcolor[HTML]{EFEFEF}\checkmark (905) & \cellcolor[HTML]{EFEFEF}\checkmark (1,169) & \cellcolor[HTML]{EFEFEF} & \cellcolor[HTML]{EFEFEF}\checkmark (1,271) & \cellcolor[HTML]{EFEFEF} & \cellcolor[HTML]{EFEFEF}\checkmark (244) & \cellcolor[HTML]{EFEFEF} & \cellcolor[HTML]{EFEFEF} & \cellcolor[HTML]{EFEFEF}1,371 \\
 & Default Function Visibility &  & \checkmark (24) &  & \checkmark (20) &  & \checkmark (2) &  &  & 29 \\
 & \cellcolor[HTML]{EFEFEF}Initializing Method without   Permission Check & \cellcolor[HTML]{EFEFEF}\checkmark (0) & \cellcolor[HTML]{EFEFEF}\checkmark (0) & \cellcolor[HTML]{EFEFEF} & \cellcolor[HTML]{EFEFEF}\checkmark (1) & \cellcolor[HTML]{EFEFEF} & \cellcolor[HTML]{EFEFEF} & \cellcolor[HTML]{EFEFEF} & \cellcolor[HTML]{EFEFEF} & \cellcolor[HTML]{EFEFEF}5 \\
 & Method \texttt{permit()} Used   for Arbitrary \texttt{from} in \texttt{transferFrom()} & \checkmark (0) & \checkmark (1) &  & \checkmark (2) &  &  &  &  & 6 \\
 & \cellcolor[HTML]{EFEFEF}Missing \texttt{msg.sender}   Check for \texttt{transferFrom()} & \cellcolor[HTML]{EFEFEF}\checkmark (0) & \cellcolor[HTML]{EFEFEF}\checkmark (0) & \cellcolor[HTML]{EFEFEF} & \cellcolor[HTML]{EFEFEF}\checkmark (2) & \cellcolor[HTML]{EFEFEF} & \cellcolor[HTML]{EFEFEF} & \cellcolor[HTML]{EFEFEF} & \cellcolor[HTML]{EFEFEF} & \cellcolor[HTML]{EFEFEF}7 \\
 & Missing Input Validation   (Call Inject) & \checkmark (0) & \checkmark (13) &  & \checkmark (12) &  & \checkmark (0) &  &  & 24 \\
 & \cellcolor[HTML]{EFEFEF}Sending Ether to Arbitrary   Destinations & \cellcolor[HTML]{EFEFEF}\checkmark (0) & \cellcolor[HTML]{EFEFEF}\checkmark (26) & \cellcolor[HTML]{EFEFEF} & \cellcolor[HTML]{EFEFEF}\checkmark (46) & \cellcolor[HTML]{EFEFEF} & \cellcolor[HTML]{EFEFEF}\checkmark (0) & \cellcolor[HTML]{EFEFEF} & \cellcolor[HTML]{EFEFEF} & \cellcolor[HTML]{EFEFEF}56 \\
 & Unprotected Contract   Destruction & \checkmark (0) & \checkmark (4) & \checkmark (13) & \checkmark (16) & \checkmark (0) & \checkmark (2) & \checkmark (1) & \checkmark (1) & 19 \\
 & \cellcolor[HTML]{EFEFEF}Unprotected Ether Withdrawal & \cellcolor[HTML]{EFEFEF}\checkmark (2) & \cellcolor[HTML]{EFEFEF}\checkmark (36) & \cellcolor[HTML]{EFEFEF}\checkmark (40) & \cellcolor[HTML]{EFEFEF}\checkmark (48) & \cellcolor[HTML]{EFEFEF}\checkmark (0) & \cellcolor[HTML]{EFEFEF}\checkmark (16) & \cellcolor[HTML]{EFEFEF} & \cellcolor[HTML]{EFEFEF} & \cellcolor[HTML]{EFEFEF}67 \\
 & Unsafe Delegatecall & \checkmark (0) & \checkmark (3) & \checkmark (1) & \checkmark (0) & \checkmark (0) & \checkmark (0) &  &  & 8 \\
 & \cellcolor[HTML]{EFEFEF}Unused Return Value & \cellcolor[HTML]{EFEFEF}\checkmark (0) & \cellcolor[HTML]{EFEFEF}\checkmark (0) & \cellcolor[HTML]{EFEFEF}\checkmark (0) & \cellcolor[HTML]{EFEFEF}\checkmark (3) & \cellcolor[HTML]{EFEFEF}\checkmark (0) & \cellcolor[HTML]{EFEFEF}\checkmark (0) & \cellcolor[HTML]{EFEFEF} & \cellcolor[HTML]{EFEFEF} & \cellcolor[HTML]{EFEFEF}6 \\
 & Usage of Public Mint or Burn & \checkmark (0) & \checkmark (1) & \checkmark (3) & \checkmark (6) &  &  &  &  & 6 \\
\multirow{-18}{*}{\textbf{Access Control   (18)}} & \cellcolor[HTML]{EFEFEF}Write to Arbitrary Storage   Location & \cellcolor[HTML]{EFEFEF}\checkmark (0) & \cellcolor[HTML]{EFEFEF}\checkmark (2) & \cellcolor[HTML]{EFEFEF}\checkmark (0) & \cellcolor[HTML]{EFEFEF}\checkmark (2) & \cellcolor[HTML]{EFEFEF} & \cellcolor[HTML]{EFEFEF}\checkmark (0) & \cellcolor[HTML]{EFEFEF} & \cellcolor[HTML]{EFEFEF} & \cellcolor[HTML]{EFEFEF}6 \\ \hline
 & Inappropriate Integer Division before Multiplication & \checkmark (0) & \checkmark (8) &  & \checkmark (2) &  & \checkmark (0) & \checkmark (1) &  & 10 \\
 & \cellcolor[HTML]{EFEFEF}Integer Overflow/Underflow & \cellcolor[HTML]{EFEFEF}\checkmark (0) & \cellcolor[HTML]{EFEFEF} & \cellcolor[HTML]{EFEFEF}\checkmark (1) & \cellcolor[HTML]{EFEFEF}\checkmark (645) & \cellcolor[HTML]{EFEFEF}\checkmark (0) & \cellcolor[HTML]{EFEFEF}\checkmark (44) & \cellcolor[HTML]{EFEFEF}\checkmark (325) & \cellcolor[HTML]{EFEFEF}\checkmark (91) & \cellcolor[HTML]{EFEFEF}1,976 \\
\multirow{-3}{*}{\textbf{Arithmetic (3)}} & Unsafe Array Length   Assignment & \checkmark (0) & \checkmark (2) & \checkmark (0) & \checkmark (4) &  &  &  &  & 6 \\ \hline
 & \cellcolor[HTML]{EFEFEF}Dangerous Usage of \texttt{block.timestamp} & \cellcolor[HTML]{EFEFEF}\checkmark (370) & \cellcolor[HTML]{EFEFEF}\checkmark (1,108) & \cellcolor[HTML]{EFEFEF} & \cellcolor[HTML]{EFEFEF}\checkmark (1,132) & \cellcolor[HTML]{EFEFEF}\checkmark (0) & \cellcolor[HTML]{EFEFEF}\checkmark (210) & \cellcolor[HTML]{EFEFEF}\checkmark (4) & \cellcolor[HTML]{EFEFEF}\checkmark (0) & \cellcolor[HTML]{EFEFEF}1,139 \\
 & Transaction Order Dependency & \checkmark (1,231) &  & \checkmark (479) & \checkmark (1,970) &  & \checkmark (0) & \checkmark (0) & \checkmark (0) & 2,687 \\
\multirow{-3}{*}{\textbf{Block Manipulation   (3)}} & \cellcolor[HTML]{EFEFEF}Weak PRNG (Pseudorandom   Number Generator) & \cellcolor[HTML]{EFEFEF}\checkmark (1) & \cellcolor[HTML]{EFEFEF}\checkmark (15) & \cellcolor[HTML]{EFEFEF}\checkmark (1) & \cellcolor[HTML]{EFEFEF}\checkmark (23) & \cellcolor[HTML]{EFEFEF}\checkmark (0) & \cellcolor[HTML]{EFEFEF} & \cellcolor[HTML]{EFEFEF} & \cellcolor[HTML]{EFEFEF} & \cellcolor[HTML]{EFEFEF}60 \\ \hline
 & Lack of Proper Signature Verification & \checkmark (0) & \checkmark (0) &  & \checkmark (1) &  &  &  &  & 1 \\
\multirow{-2}{*}{\textbf{Cryptographic (2)}} & \cellcolor[HTML]{EFEFEF}Signature Malleability & \cellcolor[HTML]{EFEFEF} & \cellcolor[HTML]{EFEFEF}\checkmark (1) & \cellcolor[HTML]{EFEFEF} & \cellcolor[HTML]{EFEFEF}\checkmark (0) & \cellcolor[HTML]{EFEFEF} & \cellcolor[HTML]{EFEFEF} & \cellcolor[HTML]{EFEFEF} & \cellcolor[HTML]{EFEFEF} & \cellcolor[HTML]{EFEFEF}1 \\ \hline
 & \texttt{transfer()} and \texttt{send()} with Hardcoded Gas   Amount & \checkmark (0) & \checkmark (0) & \checkmark (7) & \checkmark (0) &  &  &  &  & 4 \\
 & \cellcolor[HTML]{EFEFEF}Contract Could Lock Ether & \cellcolor[HTML]{EFEFEF}\checkmark (50) & \cellcolor[HTML]{EFEFEF}\checkmark (69) & \cellcolor[HTML]{EFEFEF} & \cellcolor[HTML]{EFEFEF}\checkmark (82) & \cellcolor[HTML]{EFEFEF} & \cellcolor[HTML]{EFEFEF} & \cellcolor[HTML]{EFEFEF} & \cellcolor[HTML]{EFEFEF} & \cellcolor[HTML]{EFEFEF}92 \\
 & DoS with Block Gas Limit & \checkmark (0) & \checkmark (0) &  & \checkmark (0) &  &  &  &  & 8 \\
 & \cellcolor[HTML]{EFEFEF}DoS With Failed Call & \cellcolor[HTML]{EFEFEF}\checkmark (0) & \cellcolor[HTML]{EFEFEF}\checkmark (0) & \cellcolor[HTML]{EFEFEF} & \cellcolor[HTML]{EFEFEF}\checkmark (10) & \cellcolor[HTML]{EFEFEF} & \cellcolor[HTML]{EFEFEF}\checkmark (2) & \cellcolor[HTML]{EFEFEF}\checkmark (0) & \cellcolor[HTML]{EFEFEF}\checkmark (0) & \cellcolor[HTML]{EFEFEF}15 \\
 & Force Sending Ether with  \texttt{this.balance} Check & \checkmark (0) & \checkmark (0) &  & \checkmark (0) &  &  &  &  & 1 \\
\multirow{-6}{*}{\textbf{Denial of Services   (6)}} & \cellcolor[HTML]{EFEFEF}Unsafe \texttt{send()} in the   \texttt{require()} Condition & \cellcolor[HTML]{EFEFEF}\checkmark (833) & \cellcolor[HTML]{EFEFEF}\checkmark (0) & \cellcolor[HTML]{EFEFEF}\checkmark (8) & \cellcolor[HTML]{EFEFEF}\checkmark (743) & \cellcolor[HTML]{EFEFEF}\checkmark (0) & \cellcolor[HTML]{EFEFEF}\checkmark (0) & \cellcolor[HTML]{EFEFEF} & \cellcolor[HTML]{EFEFEF} & \cellcolor[HTML]{EFEFEF}1,269 \\ \hline
 & Reentrancy Vulnerability with Negative Events & \checkmark (13) & \checkmark (111) &  & \checkmark (119) &  &  &  &  & 124 \\
 & \cellcolor[HTML]{EFEFEF}Reentrancy Vulnerability with   Transfer & \cellcolor[HTML]{EFEFEF}\checkmark (37) & \cellcolor[HTML]{EFEFEF}\checkmark (201) & \cellcolor[HTML]{EFEFEF}\checkmark (201) & \cellcolor[HTML]{EFEFEF}\checkmark (227) & \cellcolor[HTML]{EFEFEF}\checkmark (0) & \cellcolor[HTML]{EFEFEF}\checkmark (45) & \cellcolor[HTML]{EFEFEF}\checkmark (20) & \cellcolor[HTML]{EFEFEF}\checkmark (5) & \cellcolor[HTML]{EFEFEF}227 \\
 & Reentrancy Vulnerability with   Same Effect & \checkmark (69) & \checkmark (276) &  & \checkmark (297) &  &  &  &  & 331 \\
 & \cellcolor[HTML]{EFEFEF}Reentrancy Vulnerability with   Token Transfer & \cellcolor[HTML]{EFEFEF}\checkmark (83) & \cellcolor[HTML]{EFEFEF}\checkmark (450) & \cellcolor[HTML]{EFEFEF} & \cellcolor[HTML]{EFEFEF}\checkmark (513) & \cellcolor[HTML]{EFEFEF}\checkmark (0) & \cellcolor[HTML]{EFEFEF}\checkmark (53) & \cellcolor[HTML]{EFEFEF}\checkmark (27) & \cellcolor[HTML]{EFEFEF}\checkmark (27) & \cellcolor[HTML]{EFEFEF}525 \\
\multirow{-5}{*}{\textbf{Reentrancy (5)}} & Reentrancy Vulnerability   without Token Transfer & \checkmark (34) & \checkmark (190) &  & \checkmark (218) &  &  &  &  & 227 \\ \hline
 & \cellcolor[HTML]{EFEFEF}Arbitrary Function Jump via Inline Assembly{$^*$} & \cellcolor[HTML]{EFEFEF} & \cellcolor[HTML]{EFEFEF}\checkmark & \cellcolor[HTML]{EFEFEF} & \cellcolor[HTML]{EFEFEF} & \cellcolor[HTML]{EFEFEF} & \cellcolor[HTML]{EFEFEF} & \cellcolor[HTML]{EFEFEF} & \cellcolor[HTML]{EFEFEF} & \cellcolor[HTML]{EFEFEF}- \\
 & Bytes Variables Risk{$^*$} &  & \checkmark &  & \checkmark &  &  &  &  & - \\
 & \cellcolor[HTML]{EFEFEF}Dangerous Usage of   \texttt{msg.value} inside a Loop{$^*$} & \cellcolor[HTML]{EFEFEF}\checkmark & \cellcolor[HTML]{EFEFEF}\checkmark & \cellcolor[HTML]{EFEFEF} & \cellcolor[HTML]{EFEFEF}\checkmark & \cellcolor[HTML]{EFEFEF} & \cellcolor[HTML]{EFEFEF} & \cellcolor[HTML]{EFEFEF} & \cellcolor[HTML]{EFEFEF} & \cellcolor[HTML]{EFEFEF}- \\
 & Error-prone Assembly Usage & \checkmark (0) & \checkmark (0) &  & \checkmark (0) & \checkmark (0) &  &  &  & 1 \\
 & \cellcolor[HTML]{EFEFEF}Memory Manipulation & \cellcolor[HTML]{EFEFEF}\checkmark (0) & \cellcolor[HTML]{EFEFEF}\checkmark (0) & \cellcolor[HTML]{EFEFEF} & \cellcolor[HTML]{EFEFEF}\checkmark (1) & \cellcolor[HTML]{EFEFEF} & \cellcolor[HTML]{EFEFEF} & \cellcolor[HTML]{EFEFEF} & \cellcolor[HTML]{EFEFEF} & \cellcolor[HTML]{EFEFEF}2 \\
 & Modifying Storage Array by   Value{$^*$} & \checkmark & \checkmark &  & \checkmark &  &  &  &  & - \\
 & \cellcolor[HTML]{EFEFEF}Payable Functions Using   \texttt{delegatecall} inside a Loop & \cellcolor[HTML]{EFEFEF}\checkmark (0) & \cellcolor[HTML]{EFEFEF}\checkmark (0) & \cellcolor[HTML]{EFEFEF} & \cellcolor[HTML]{EFEFEF}\checkmark (1) & \cellcolor[HTML]{EFEFEF} & \cellcolor[HTML]{EFEFEF} & \cellcolor[HTML]{EFEFEF} & \cellcolor[HTML]{EFEFEF} & \cellcolor[HTML]{EFEFEF}2 \\
\multirow{-8}{*}{\textbf{Storage \&   Memory (8)}} & Uninitialized variable & \checkmark (0) & \checkmark (23) & \checkmark (0) & \checkmark (31) &  &  &  &  & 35 \\ \hline
\hline
\multicolumn{2}{c|}{\textbf{\# SWC}} & {16} & {17} & {8} & {\textbf{32}} & {8} & {14} & {8} & {6} & - \\\hline
\multicolumn{2}{c|}{\textbf{\# Supported}} & {37} & {43} & {14} & {\textbf{44}} & {11} & {17} & {8} & {7} & -\\\hline
\end{tabular}
}
\end{table}


\subsection{Dataset Collection}\label{datasets}
\subsubsection{Benchmark Construction}\label{benchmark}
To create a diverse benchmark covering as many vulnerability types as possible from our taxonomy for a thorough SAST tools evaluation, we defined three criteria for collecting datasets as follows: 

\noindent\textit{\textbf{Criterion \#1 (Solidity smart contracts and availability):}} 
The benchmark must consist of open-source Solidity smart contracts. Since our evaluation focuses on tools specifically designed for Solidity, it is essential to collect appropriate language datasets that are readily available. We identified 19 open-source datasets that fulfill this requirement.

\noindent\textit{\textbf{Criterion \#2 (Popularity or peer-reviewed):}} 
The benchmark should be either widely used or peer-reviewed. This ensures that the dataset has been recognized and applied in related research papers, demonstrating its relevance and validity in the research community. {We thereby included 7 benchmarks
including~\cite{issta20-bug_injection, ren_2021, icse23, jiachi-define, not-so, sb-curated, sb-wild}. }

\noindent\textit{\textbf{Criterion \#3 (Labeled vulnerability types):}} This criterion covers two aspects: \ding{172} The benchmark should primarily focus on vulnerabilities, and \ding{173} the ground truth should be labeled at least at the file level. 
{As we aim to gather high-quality datasets and further label them at the function level, we excluded 3 benchmarks with no ground truth or labeled only at the project level, such as~\cite{icse23, jiachi-define, sb-wild}. 
Finally, we obtained 4 available benchmarks~\cite{issta20-bug_injection, ren_2021, not-so, sb-curated}. }

Furthermore, we tried to collect more real-world vulnerable smart contracts. To this end, we focused on selecting representative BNB projects from BscScan~\cite{bscscan}. To obtain a representative sample, we first gathered the top 3,000 BNB projects ranked by their \textit{Market Capitalization (Market Cap)} and \textit{Liquidity Value (Liquidity)} as indicators. 
\textit{Market Cap} reflects a project's total market value, indicating its prominence and adoption. \textit{Liquidity} reflects the ease of trading a project's token, with higher values suggesting more accessible and stable markets. By selecting projects with high \textit{Market Cap} and \textit{Liquidity}, we ensured our dataset includes prominent and widely-used BNB contracts. After filtering out unavailable contracts, we obtained 2,941 addresses corresponding to 2,941 unique projects. We eventually got 2,941 representative BNB projects containing 8,249 smart contract files, focusing specifically on those with a \textit{Market Cap} greater than \$3,000 and the average \textit{Liquidity} of \$1,632,386, respectively. 
After this process, we collaborated with our industrial partner by engaging 3 security audit experts to construct the ground truth. \revised{We identified 120 vulnerabilities affecting 113 contracts from these BNB projects (function level). 
}

To remove duplicates from all five datasets within our benchmark suite, we checked the MD5 checksums of the contract files after eliminating blank lines and comments, a method commonly used in related studies~\cite{icse20-large, kr-study}. {In the end, we obtained 8,981 unique contract files. Among these, there are 788 vulnerable contract files containing 10,394 vulnerabilities at the function level (refer to~\Cref{tab:taxonomy} and Table~\ref{tab: benchmarks} for details). }
\smallskip

\begin{minipage}{\textwidth}
  \begin{minipage}[]{0.55\textwidth}
    \centering
    \captionof{table}{The composition of our benchmark. MI denotes manually injected vulnerabilities and RW for real-world vulnerabilities.}
    \label{tab: benchmarks}
    \setlength{\tabcolsep}{1mm}{
        \resizebox{0.98\columnwidth}{!}{
            \begin{tabular}{@{}ccccr@{}}
                \toprule
                \multirow{2}{*}{\textbf{Datasets}} & \multirow{2}{*}{\textbf{\# Files}} & \multicolumn{2}{c}{\textbf{\# Vulnerabilities}} & \multicolumn{1}{c}{\multirow{2}{*}{\textbf{Source}}} \\ \cmidrule(lr){3-4}
                                                         &       & \textbf{MI}  & \textbf{RW}  &                                       \\ \midrule
                \textbf{Not So Smart Contracts}          & 18    & \XSolidBrush & 18           & Trail of Bits~\cite{not-so}           \\
                \rowcolor[HTML]{EFEFEF} 
                \textbf{Smartbugs-Curated}               & 143    & \XSolidBrush & 257         & ICSE'20~\cite{sb-curated}           \\
                \textbf{SolidiFI Benchmark}              & 300   & 9,606        & \XSolidBrush & ISSTA'20~\cite{issta20-bug_injection} \\
                \rowcolor[HTML]{EFEFEF} 
                \textbf{Smart Contract Benchmark Suites} & 214   & \XSolidBrush & 393          & ISSTA'21~\cite{ren_2021}              \\
                {\textbf{BNB Benchmark}}  & 113   & \XSolidBrush   & 120        & BscScan~\cite{bscscan}               \\
                \bottomrule
            \end{tabular}
        }
    }
  \end{minipage}
\hspace{5pt}
  \begin{minipage}[]{0.35\textwidth}
    \centering
		\includegraphics[width=0.98\textwidth]{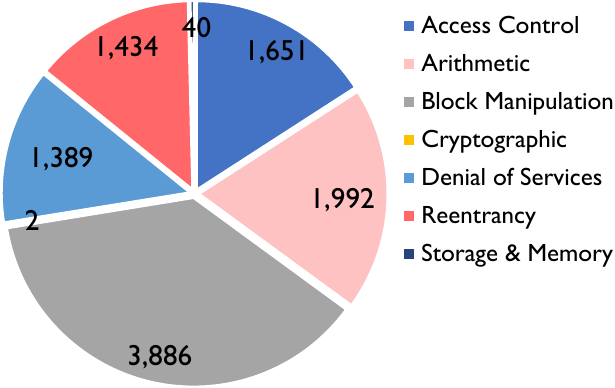}
		\captionof{figure}{Distribution of vulnerabilities by category in our benchmark.}
		\label{fig:distribution}
  \end{minipage}
\end{minipage}
\smallskip

\subsubsection{{Mapping {Vulnerability Identifiers} and Benchmark Data to Our Taxonomy}}\label{mapping}

To automatically evaluate these tools, one challenge we faced was that SAST tools use different vulnerability identifiers for their supported vulnerability types. For example, Mythril employs SWC entries in its reported issues, while others introduce their own vulnerability identifiers. This disparity makes it difficult to automatically determine whether a SAST tool detected a specific vulnerability type. {Hence, we tried to map their vulnerability identifiers to the unified vulnerability types in our taxonomy. This required a thorough review of the vulnerability identifiers, descriptions, and associated source code provided by each tool.}
Furthermore, although the four datasets we collected have already classified the vulnerability data, the vulnerability categories and types differ since they use different taxonomies. Additionally, some of them~\cite{not-so,ren_2021} specify the vulnerability location at the file level. To conduct a much fairer evaluation, we relabeled and remapped these vulnerability ground truths to our proposed vulnerability taxonomy at the function level. 
To ensure accuracy and minimize potential noise during this phase, we involved three security auditing experts for manual cross-validation of the data. In total, it took us 7.5 person-months to relabel and map the entire benchmark. An additional 3.5 person-months were spent on mapping {vulnerability identifiers} and cross-validating the mapping results (details in \S~\ref{internal_threat}). 


As displayed in Table~\ref{tab:taxonomy} and~\Cref{fig:distribution}, our benchmark has a nearly complete vulnerability type coverage (88.9\%, 40/45) on the proposed vulnerability taxonomy. 
Importantly, we attempted to incorporate test cases for the missing five vulnerability types from real-world industrial BNB projects; however, we were unable to locate any suitable examples. Our industry auditing experts confirm that the lack of these vulnerability types in our benchmark is due to their rare occurrence in practical projects. 
This observation further underscores the benchmark's alignment with the contemporary landscape of smart contract vulnerabilities.
\revised{
Meanwhile, the distribution of vulnerability samples presented in~\Cref{tab:taxonomy} reveals notable skew, with certain types represented by a significantly larger number of samples. This was primarily attributed to the inclusion of the SolidiFI Benchmark~\cite{issta20-bug_injection}. 
Despite the observed skew, the decision to include the SolidiFI Benchmark was driven by the lack of available alternatives. 
This made SolidiFI Benchmark an essential choice for our study, ensuring a broad and valuable dataset for our analysis 
.
} 
To the best of our knowledge, it is \textbf{the largest smart contract vulnerability benchmark \revised{in Solidity} (ground truth) at the function level.}

\subsection{Research Questions}\label{rq}
\revised{Based on the selected tools, proposed taxonomy, and our benchmark, we aim to answer the following research questions (RQs):}

\subsubsection{{RQ1: Coverage Analysis.}} \revised{To what extent do existing SAST tools support different vulnerability types?}

This RQ first explores the actual coverage of these generalized SAST tools across the vulnerability types defined in our taxonomy (\S~\ref{taxonomy_construction}). We assess the tools' detection capabilities to understand their comprehensiveness in identifying a wide spectrum of vulnerabilities. To this end, we leverage the previously established mapping of vulnerability identifiers to our taxonomy in \S~\ref{mapping}, aiming to quantify the effectiveness of these tools in safeguarding smart contracts against diverse security threats. Additionally, we provide the number of SWC entries to which the vulnerability identifiers can be mapped (i.e., \# SWC in \Cref{tab:taxonomy}). 
To quantitatively evaluate the scope of vulnerabilities supported by each tool against our taxonomy, we define Coverage ($= \frac{\# \text{ Supported vulnerability types}}{\# \text{ Vulnerability types in our taxonomy}}$) to measure \textit{the proportion of vulnerability types within our taxonomy that a tool claim to identify}. 

\subsubsection{{RQ2: Effectiveness Analysis.}} How effective are these SAST tools in detecting vulnerabilities on our benchmark?

This RQ is addressed from two angles: \textit{1)} The capability of current tools to effectively analyze our benchmark, and \textit{2)} the number of vulnerabilities identified within our benchmark by these tools.
For the initial aspect, we conducted a comprehensive review of the outcomes of the analyses, categorizing them into three distinct groups based on their performance: Successful scans, Scans that failed due to timeouts, and Scans that failed due to compilation issues.
Regarding the second aspect, we enumerated the vulnerabilities detected across various types within our taxonomy. 
Subsequently, we calculated Recall (\(\frac{TP}{TP+FN}\)), Precision (\(\frac{TP}{TP+FP}\)), and F1-score (\(\frac{2 \times Recall \times Precision}{Recall + Precision}\)), to evaluate the effectiveness of these tools at function level. 

\revised{
All experiments in this study were performed on a server equipped with 80 vCPUs (Intel$^\circledR$ Xeon$^\circledR$ Gold 6248 CPU @ 2.50 GHz ×2) and 188G of RAMs, which uses GNU/Linux Ubuntu 18.04 (64-bit) as the host operating system. To automate the execution of these tools, we extended the SmartBugs~\cite{smartbugs} framework and enabled all detectors capable of identifying security vulnerabilities, as well as leveraging the recommended runtime parameters in~\cite{ren_2021}, in which they proposed a unified standard to eliminate the bias in the assessment process.
}
\subsubsection{RQ3: Consistency Analysis.} This research question focuses on two consistency analyses: \textit{1)} Are the detection results consistent among these tools in terms of the detected vulnerability categories? \textit{2)} How effective are these SAST tools when combining their detecting results? 

\revised{
Building upon the empirical findings from RQ2, this research question aims to dissect the disparities among tools in their vulnerability detection. By meticulously analyzing both the commonalities and unique attributes of these tools' detection capabilities, we aim to shed light on their distinct strengths and potential shortcomings. This analysis focuses on: 
\ding{172} How these tools vary in detection rates across vulnerability categories. 
\ding{173} The potential boost in detection when combining tools, leveraging their individual strengths.
}
\subsubsection{{RQ4: Efficiency Analysis.}} How efficient are these SAST tools to perform an analysis?

\revised{
To thoroughly assess the efficiency of various SAST tools on real-world smart contracts, we analyze all 8,981 contract files collected in \S~\ref{benchmark}. For a robust and reliable evaluation under diverse conditions, we consider the efficiency based on all 8,981 contract files regardless of the success or failure of the analysis. 
To ensure robustness and consider potential infrastructure variability, we performed each performance measurement three times for each tool. The reported results represent the average of these trials. 
}

\section{Comparison and Evaluation}

\subsection{{RQ1: Coverage Analysis}}\label{sec:rq1-results}

\revised{As depicted in Table~\ref{tab:taxonomy}, the vulnerability coverage varies among tools, despite they are generalized-focused tools. 
A noteworthy observation is the significant discrepancy in coverage between the commercial tool CSA, which supports nearly all types except for ``Arbitrary Function Jump via Inline Assembly'' (achieving 97.78\%, 44/45), and the open-source tools. 
Among these, Slither stands out with the second-highest coverage at 95.56\% (43/45), followed by Securify2 at 82.22\% (37/45). This variation, particularly with Oyente showing lower coverage, highlights the impact of tool maturity on the breadth of supported vulnerability types, with older tools like Oyente lagging in updates for newer vulnerabilities. 
Meanwhile, it further illuminates areas requiring attention from tool developers. Specifically, the \textit{Storage \& Memory} and \textit{Cryptographic} categories present lower coverage across the board, indicating critical directions for future enhancements. 
Interestingly, Slither exhibits comprehensive coverage in these categories, positioning it as a benchmark for vulnerability detection in these domains. 
}
\begin{tcolorbox}[size=title,opacityfill=0.1, boxsep=0mm]
\noindent\textbf{Finding 1:} CSA, Slither, and Securify2 demonstrate the highest coverage across multiple categories. Slither excels in the \textit{Reentrancy} and \textit{Storage \& Memory} categories, while CSA stands out in all categories except for \textit{Storage \& Memory}. 
\end{tcolorbox}

\revised{
Notably, Slither ranks second in coverage yet misses two prevalent vulnerability types. Specifically, \ding{172} in the \textit{Arithmetic} category, Slither fails to support \textit{Integer Overflow/Underflow}, which are frequently identified in smart contracts~\cite{so2020verismart}. 
While the Solidity compiler has built-in checks for these issues~\cite{Solidity0.8.0} from version \texttt{v0.8.0}~\cite{Solidity0.8.0}, the inclusion of the \texttt{unchecked} keyword allows developers to bypass these protections for the sake of gas optimization. This bypass reintroduces potential overflow/underflow risks that the compiler's checks aimed to eliminate. 
It shows a crucial gap in Slither's analysis capabilities. Users should be aware of this when seeking overflow/underflow analysis, especially where compiler protections might be bypassed. 
\ding{173} Moreover, Slither lacks rules for detecting Transaction Order Dependence (TOD) vulnerabilities. These vulnerabilities are a common concern~\cite{wang2019detecting,SWC114Sm97:online,zhang2022front} and are supported by all other tools except for Manticore. The absence of TOD support in them (both developed by Trail of Bits) suggests a potential oversight and a critical area for improvement. 
}

\revised{Other tools, such as Mythril, Manticore, and Osiris, display selective coverage, prioritizing specific vulnerability categories. For instance, Mythril concentrates on \textit{Access Control}}. 
\revised{Meanwhile, our analysis of tool support revealed that Manticore is equipped with three specific vulnerability identifiers for ``Unsafe \texttt{Delegatecall}'' within the \textit{Access Control} category. This might suggest a targeted emphasis by Manticore on detecting this particular type of vulnerability. }
Moreover, despite having a smaller set of vulnerability identifiers, Osiris prioritizes the \textit{Arithmetic} category with five out of its nine vulnerability identifiers targeting it. It covers four distinct vulnerability types within this category, demonstrating its granular focus despite limited overall coverage. However, Oyente and Osiris cover only four categories, potentially limiting their ability to detect a variety of vulnerabilities across categories. 

\begin{tcolorbox}[size=title,opacityfill=0.1, boxsep=0mm]
\noindent\textbf{Finding 2:} {Other tools, such as Mythril, Manticore, and Osiris, show selective coverage, focusing on specific vulnerability categories. While such focused coverage can lead to detailed vulnerability identification in those categories, it also results in limited overall coverage. 
}
\end{tcolorbox}

\subsection{RQ2: Effectiveness Analysis}\label{sec:rq2-results}

\begin{table}
\centering
\caption{Analysis status overview for each SAST tool. 
\revised{{\# Success} and {\# Failure} indicate the number of contracts where tools scan successfully and unsuccessfully, respectively. \# Timeout indicates the number of contracts where the tool failed due to exceeding a time limit. 
{\# Compilation} denotes the number of contracts that could not be scanned due to compilation errors. }
}\label{tab:analysis_info}
\resizebox{0.65\columnwidth}{!}{
\begin{tabular}{@{}ccccc@{}}
\toprule
\multirow{2}{*}{\textbf{Tool}} & \multirow{2}{*}{\textbf{\# Success}} & \multicolumn{3}{c}{\textbf{\# Failure}} \\ \cmidrule(l){3-5} 
                    &                     & \textbf{\# Timeout} & \textbf{\revised{\# Compilation}} & \textbf{\# Total} \\ \midrule
\textbf{Securify2}  & 247 (31.35\%)        & 38                 & \textbf{503}       & 541              \\
\textbf{Slither}    & 767 (97.34\%)        & 2                  & 19                 & 21               \\
\textbf{SmartCheck} & \textbf{788 (100.00\%)} & 0                  & 0                  & 0                \\
\textbf{CSA}        & 772 (97.97\%)        & 0                  & 16                 & 16               \\
\textbf{Manticore}  & 112 (14.21\%)        & \textbf{626}       & 50                 & \textbf{676}     \\
\textbf{Mythril}    & 590 (74.87\%)        & 193                & 0                  & 193              \\
\textbf{Osiris}     & 451 (57.23\%)        & 35                 & 302                & 337              \\
\textbf{Oyente}     & 499 (63.32\%)        & 57                 & 232                & 289              \\ \bottomrule
\end{tabular}
}
\end{table}

\begin{figure*}[]
    \includegraphics[width=.5\linewidth]{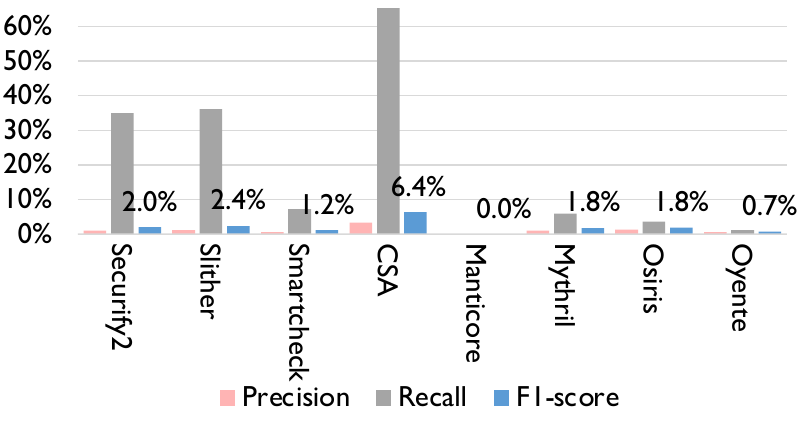}
	\caption{Overall effectiveness of each tool on our benchmark.}
    \label{fig:overall}
\end{figure*}

As shown in Table~\ref{tab:analysis_info}, we first investigated the analysis status due to numerous cases where tools failed to analyze contracts or generate reports. 
We observed that existing tools primarily analyze smart contracts by first compiling them, except for SmartCheck. However, this approach often leads to compilation issues, causing the analyzers to fail in their analysis. 
For instance, Securify2 relies on the compiler version of contracts under examination to generate its Abstract Syntax Tree (AST), but it is limited to compiler versions between \texttt{v0.5.*} and \texttt{v0.6.*}. As a result, Securify2 fails to analyze contracts developed using other compiler versions. 
Oyente faces challenges when analyzing contracts due to its limited support for compiler versions up to \texttt{v0.4.19}. 
Consequently, Oyente failed to analyze 232 (29.44\%) contracts that were developed using higher \texttt{solc} versions with different AST structures~\cite{oyente-version}. 
Osiris, built on the foundation of Oyente, encountered a ``compilation failed'' error on 38.32\% (302) contracts as it only supports the Solidity compiler \texttt{v0.4.21}. This limitation causes Osiris to struggle with contracts written in Solidity versions beyond \texttt{v0.4.21}, triggering compilation failures and preventing further vulnerability detection, rendering Osiris less efficient compared to other analyzers. 
Interestingly, Manticore faced difficulties in analyzing 626 out of 788 contract files. More specifically, it encountered a timeout in 79.44\% (626/788) contracts and demanded substantial memory resources in our experimental system. Additionally, Manticore failed to analyze 50 contracts and generated exception errors due to compilation issues. 
To further investigate this issue, we focused on the SolidiFI Benchmark, where Manticore failed to analyze all the 300 contracts: 40 due to compilation issues and 260 due to timeout. For the 260 contracts that encountered timeouts, we removed the timeout constraint and reran Manticore to evaluate its detection capacity. 
\revised{However, we observed that over three weeks, Manticore successfully analyzed only 36 out of 260 contracts without yielding new findings. This underscores the practicality of our one-hour limit by demonstrating its sufficiency even for the
most demanding analyses. }
Additionally, upon further exploration, we found that 44.25\% (50/113) BNB projects encountered compilation failure when using Securify2, Manticore, and Oyente. 
We further investigated that tools except for CSA, Slither, and SmartCheck only support single-file analysis, resulting in them having no capacity to scan multiple files under a project and causing compilation failure, especially when these projects involve importing other files. 

\begin{tcolorbox}[size=title,opacityfill=0.1, boxsep=0mm]
            \noindent \textbf{Finding 3:}
            {Some analyzers such as Oyente, Osiris, and Securify2 run deeply tied to the Solidity compiler version, which can cause analyzing failure when trying to scan smart contracts written in different versions.}
\end{tcolorbox}

\revised{
As illustrated in~\Cref{fig:overall}, the tools exhibit a wide range of effectiveness in vulnerability detection, with many performing below expectations. CSA stands out in terms of F1-score, with its metrics revealing a Recall of 72.00\%, Precision of 3.35\%, and an F1-score of 6.40\%. Slither follows with an F1-score of 2.38\%, Recall of 36.18\%, and Precision of 1.23\%. Other tools, such as Manticore, show disappointing results, failing to detect any vulnerabilities in our benchmark, resulting in an F1-score of 0\% due to numerous analysis failures. Interestingly, static analysis (SA) tools generally achieve better Recall, whereas symbolic execution (SE) tools tend to offer superior Precision.
}

Regarding SA tools, Intermediate Representation (IR) has become a cornerstone to translate high-level source code to actionable, low-level insights. All four SA tools studied (i.e., Securify2, Slither, SmartCheck, and CSA) employ IR, yet their quality varies. 
Both Slither and Securify2 harness IR by compiling the code, allowing them to conduct control flow graph (CFG) and data flow graph (DFG) analyses with enhanced precision~\cite{slither_ir,securify2_ir}. This meticulous compilation-based approach provides a robust depiction of code behavior during execution. 
In contrast, SmartCheck adopts ANTLR~\cite{parr1995antlr} to directly transform Solidity code into an XML-based IR~\cite{smartcheck_pub}. This non-compilation approach enhances its analysis success rate, as it sidesteps potential compilation challenges mentioned before. However, the trade-off is its abstracted IR, which may lack the granularity to capture all code intricacies. This is underscored by SmartCheck's performance on the SolidiFI Benchmark: while it detects an average of $1,490$ vulnerabilities per contract, its precision stands at a mere $0.65\%$. 
This highlights the balance between the detailed methods of Slither and Securify2 versus the more generalized but potentially less precise approach of SmartCheck. 

Generally, these SA tools struggle with high false positives, resulting in an average F1-score below 10\%. 
An intriguing relationship emerges between false negatives and false positives: 
The four SA tools (Securify2, Slither, SmartCheck, and CSA) focused on source code and attained elevated Recall, which is attributed to their inherent technical constraints, which prioritize achieving a \textit{sound} detection by compromising the \textit{completeness}~\cite{boolos2002computability}. However, it results in a trade-off, manifesting in diminished Precision and an increased number of false positives. 

\revised{
Take Slither as an example, despite utilizing advanced detection techniques involving data-flow and control-flow analysis on the \texttt{SlithIR}, it still suffers from unexpected false positives. A notable instance is its detection logic for ``Weak PRNG (Pseudorandom Number Generator)'', which refers to compromised PRNG due to predictable or manipulable sources like using \texttt{blockhash}, with the primary concern being the security risk from predictable outcomes. 
However, the detector implementation within Slither~\cite{slither_prng} utilizes Static Single Assignment (SSA) form to identify potentially risky uses of the modulus operator when its left operand depends on \texttt{block.timestamp}, \texttt{now}, or \texttt{blockhash}. Despite this sophisticated approach, it frequently misidentifies benign code as vulnerable (resulting in false positives) in practice. 
For instance, as shown in~\Cref{lst:prng_func}, the seven instances reported as ``Weak PRNG'' vulnerabilities in the PESA Token~\cite{BNB_case} were all false positives (L2, L5, L8, L14, L17, L20, and L23). However, the code demonstrates benign applications of the modulus operator for various logical operations, such as determining leap years (L1-L12) or extracting time components from a timestamp (L13-L24). 
This false positive originates from the simplistic approach of the detection heuristic for this vulnerability pattern, which lacks a nuanced understanding necessary to differentiate between the use of the modulus operator in random number generation and its benign use in straightforward arithmetic operations. 
}

\begin{lstlisting}[language=Solidity, caption={Examples of benign modulus operator usage incorrectly flagged as \texttt{Weak PRNG} vulnerabilities by Slither in the PESA Token~\cite{BNB_case}.}, label={lst:prng_func}, float=t]
function isLeapYear(uint16 year) public pure returns (bool) {
    if (year % 4 != 0) { // FP-1
            return false;  
    }
    if (year % 100 != 0) { // FP-2
            return true;
    }
    if (year % 400 != 0) { // FP-3
            return false;
    }
    return true;
}
function getHour(uint timestamp) public pure returns (uint8) {
        return uint8((timestamp / 60 / 60) % 24); // FP-4
}
function getMinute(uint timestamp) public pure returns (uint8) {
        return uint8((timestamp / 60) % 60); // FP-5
}
function getSecond(uint timestamp) public pure returns (uint8) {
        return uint8(timestamp % 60); // FP-6
}
function getWeekday(uint timestamp) public pure returns (uint8) {
        return uint8((timestamp / DAY_IN_SECONDS + 4) % 7); // FP-7
}
\end{lstlisting}

In contrast, SE tools including Mythril, Oyente, and Osiris, which analyze bytecode, tend to be more precise when compared with these SA tools. 
The enhanced precision arises from the fact that SE simulates specific code execution paths. However, while this offers a more meticulous analysis, it can sometimes overlook vulnerabilities outside the simulated paths, particularly when utilizing Depth-First Search (DFS) within constrained timeframes.

\begin{tcolorbox}[size=title,opacityfill=0.1, boxsep=0mm]
            \noindent \textbf{Finding 4:}
            {SAST tool effectiveness varies notably. CSA and Slither lead with F1-scores of 6.40\% and 2.38\%, while Oyente and Manticore underperform. The quality of IR is crucial: detailed IRs enhance accuracy, but SmartCheck's approach leads to high false positives. SA (using source code analysis) boosts Recall, while SE (using bytecode analysis) prioritizes Precision.}
\end{tcolorbox}

\subsection{RQ3: Consistency Analysis}\label{consistency}

\noindent\textbf{\textit{Detection consistency across vulnerability categories.}}
{Generally, these tools exhibited weak detection capabilities across all categories, which can be inferred from the results in RQ1. 
Interestingly, we still observed a significant inconsistency in the detection capabilities of SAST tools across various vulnerability categories, as displayed in~\Cref{tab:category_result}. For instance, these tools exhibited more effective detection rates for \textit{Access Control} (30.31\%) and \textit{Reentrancy} (28.03\%) vulnerabilities when compared to other categories. 

}
\begin{lstlisting}[language=Solidity, caption={Detection logic for \textit{unrestricted-ether-flow} within Securify2.}, label={lst:securify2_dos}, float=t]
// Identifying relevant contexts via external calls
applicableInContext(callCtx) :- externalCall(call), ctxProvider.elementInContext(callCtx, call, _).
// Compliance check: Ether amount associated with a call is zero
compliantInContext(callCtx, "") :-
    applicableInContext(callCtx), ctxProvider.elementInContext(callCtx, call, context),
    callValue(call, amount),
    valueOf([amount, context], "0").
// Compliance check: External call depends on the transaction's sender
compliantInContext(callCtx, "") :-
    applicableInContext(callCtx),
    someSender(senderCtx),
    (
        infoflow.instrMustDependOn(callCtx, senderCtx)
    ).
// Violation rule: Non-compliant conditions, indicating potential unrestricted ether flows
violationInContext(callCtx, "") :-
    applicableInContext(callCtx), ctxProvider.elementInContext(callCtx, call, context),
    callValue(call, value),
    !zeroValue([value, context]),
    !maybeCompliantInContext(callCtx).
\end{lstlisting}




            
            

In contrast, the tools demonstrated weaker effectiveness in \textit{Storage \& Memory} (1.88\%), \textit{Arithmetic} (7.05\%), and \textit{Cryptographic} (12.50\%) categories. Notably, while all tools support \textit{Arithmetic} vulnerabilities, they detected no more than ${40\%}$ of them. 
{The observed inconsistency also sheds light on the distinct methodologies and detection focuses of different SAST tools. 
\revised{For example, while most tools detected an average of 16.23\% \textit{Denial of Service} vulnerabilities, Securify2 single-handedly detected 64\% of the 1,389 vulnerabilities}. This difference suggests Securify2's unique approach. It combines both ``compliance'' and ``violation'' patterns for formal verification instead of relying solely on known sensitive signatures or heuristics. 
A prime example is its detection pattern for ``\texttt{transfer()} and \texttt{send()} with Hardcoded Gas Amount''. As depicted in~\Cref{lst:securify2_dos}, Securify2 first identifies contexts where external calls are made (L2). For compliance, it verifies if the ether amount associated with a call is zero (L4-L7) or if the external call explicitly depends on the transaction's sender (i.e., \texttt{msg.sender}) (L10-L15). Finally, violations are flagged when these compliant conditions are not met, signaling potential unrestricted ether flows (L18-L22).
However, the inherent drawback of Securify2 relying on hard-coded patterns, such as \textit{mul-after-div.dl}~\cite{securify2_div_mul}, becomes evident when faced with intricate arithmetic issues in real-world scenarios, especially in DeFi apps where diverse factors intertwine, leading to potential miscalculations or undetected vulnerabilities.}

\begin{table}[]
\centering
\caption{\revised{The number of detected vulnerabilities in each category.
For each vulnerability category: AC: \textit{Access Control}, AR: \textit{Arithmetic}, BM: \textit{Block Manipulation}, CR: \textit{Cryptographic}, DoS: \textit{Denial of Services}, RE: \textit{Reentrancy}, S\&M: \textit{Storage \& Memory}. }
}
\label{tab:category_result}
\resizebox{0.8\columnwidth}{!}{
\begin{tabular}{@{}crrrrrrrr@{}}
\toprule
\multicolumn{1}{c}{\textbf{Tool}}                      & \multicolumn{1}{c}{\textbf{AC}}      &  \multicolumn{1}{c}{\textbf{AR}}    &  \multicolumn{1}{c}{\textbf{BM}}   &  \multicolumn{1}{c}{\textbf{CR}}  &  \multicolumn{1}{c}{\textbf{DoS}} & \multicolumn{1}{c}{\textbf{RE}}    & \multicolumn{1}{c}{\textbf{S\&M}}  \\ \midrule
\textbf{Securify2}  & 909\mybar{55}   & 0\mybar{0}    & 1,602\mybar{41}  & 0\mybar{0}   & 883\mybar{64} & 236\mybar{16}  & 0\mybar{0}  \\
\textbf{Slither}    & 1,306\mybar{79} & 10\mybar{1}   & 1,123\mybar{29}  & 1\mybar{50}   & 69\mybar{4}    & 1,228\mybar{86} & 3\mybar{8} \\
\textbf{SmartCheck} & 57\mybar{21}   & 1\mybar{0}    & 480\mybar{12}   & 0\mybar{0}   & 15\mybar{1}    & 201\mybar{14}  & 0\mybar{0}  \\
\textbf{CSA}        & 1,465\mybar{89}   & 651\mybar{32} & 3,125\mybar{80} & 1\mybar{50}  & 835\mybar{60}    & 1,374\mybar{96}   & 33\mybar{8} \\
\textbf{Manticore}  & 0\mybar{0}      & 0\mybar{0}    & 0\mybar{0}      & 0\mybar{0}   & 0\mybar{0}     & 0\mybar{0}     & 0\mybar{0}   \\
\textbf{Mythril}    & 264\mybar{16}   & 44\mybar{2}   & 210\mybar{1}    & 0\mybar{0}   & 2\mybar{0}     & 98\mybar{7}  & 0\mybar{0}   \\
\textbf{Osiris}     & 1\mybar{0}      & 326\mybar{16}  & 4\mybar{0}      & 0\mybar{0}   & 0\mybar{0}     & 47\mybar{3}    & 0\mybar{0}   \\
\textbf{Oyente}     & 1\mybar{0}      & 91\mybar{5}   & 0\mybar{0}      & 0\mybar{0}   & 0\mybar{0}     & 32\mybar{2}    & 0\mybar{0}   \\ \midrule
{\textbf{\# Vulns in our benchmark}} & \multicolumn{1}{c}{1,651}             & \multicolumn{1}{c}{1,992}           & \multicolumn{1}{c}{3,886}          & \multicolumn{1}{c}{2}            & \multicolumn{1}{c}{1,389}         & \multicolumn{1}{c}{1,434}           & \multicolumn{1}{c}{40}           \\ \bottomrule
\end{tabular}
}
\end{table}


Meanwhile, CSA and Slither led in detecting \textit{Reentrancy} vulnerabilities (95.82\% and 85.63\%, respectively). We found that Slither employs a heuristic-based method to detect potential reentrancy vulnerabilities in smart contracts. \revised{By integrating 7 detectors for reentrancy detection~\cite{slither_ree,website-rq2}, it systematically traverses the CFG of each implemented function, searching for patterns that could indicate reentrancy, particularly scenarios where state variables are written after an external call.}
{However, Slither still has its limitations in detecting this category, especially when analyzing contracts involving state variables. If the callee is a state variable controlled only by an administrator, Slither might mistakenly report a non-existent vulnerability (false positive). Conversely, if the callee is a state variable that can be influenced by anyone, Slither may overlook the vulnerability (false negative).}

{
It is crucial for developers and researchers to recognize these strengths and shortcomings. This comprehension can help guide further advancements in SA tools, ensuring that they continue to evolve and offer robust defenses against potential smart contract vulnerabilities.
}

\begin{tcolorbox}[size=title,opacityfill=0.1, boxsep=0mm]
    \noindent \textbf{Finding 5:}
    {
    There is a notable difference in detection rates across vulnerability categories. 
    Interestingly, they identified more \textit{Access Control} and \textit{Reentrancy} vulnerabilities when compared to those belonging to \textit{Storage \& Memory}, \textit{Arithmetic}, and \textit{Cryptographic} categories.
    This inconsistency underscores their unique detection methodologies and the inherent challenges in improving the coverage for diverse vulnerability types.
    }
\end{tcolorbox}

\noindent\textbf{\textit{{The combination of tools.}}} 
Since none of the single tools performs well on our benchmark and each tool has different focuses, we tried to analyze the effectiveness improvement by tool combination. 

\noindent\textit{\ding{172} An inclusive combination strategy:}
Here, we selected and combined the top-performing tools in terms of Recall: CSA (72.00\%), Slither (36.18\%), and Securify2 (34.92\%) (see~\Cref{fig:combination}). A vulnerability is thereby considered found if at least one tool from the respective group was able to detect it. Interestingly, our findings demonstrate that CSA and Securify2 effectively complement each other in vulnerability detection. Specifically, as shown in~\Cref{fig:combination} (b), \revised{they identified 91.5\% (9,513/10,394) compared to the best-performing single tool, CSA (72.00\%). 
However, this improvement comes at the cost of marking 36.77 percentage points more functions as potentially vulnerable (271,794 reported issues in total). }

\begin{tcolorbox}[size=title,opacityfill=0.1,boxsep=0mm]
    \textbf{Finding 6:} {
    \revised{Applying an \textit{inclusive combination} strategy, where a vulnerability is considered found if detected by any tool, substantially raises Recall to 91.5\%, albeit at the cost of decreasing Precision to 2.67\%. }
    }
\end{tcolorbox}

\noindent\textit{\ding{173} Majority voting strategy:} 
The low Precision of single tools motivated us to explore consensus-based approaches for potentially improving Precision. 
To this end, we adopted the ``majority voting''~\cite{majority-voting} strategy, excluding Manticore due to its inability to detect any vulnerabilities, and thus considering the consensus among the remaining seven tools. 
This approach yielded a noteworthy increase in Precision to 36.10\% (717/1,986), with a corresponding improvement in the F1-score to 11.58\%. 
Compared to the single tool with the highest Precision, CSA (3.35\%), this consensus-based approach significantly improved Precision. However, Recall was reduced to 6.90\%, which underlines the trade-off between Precision and Recall and the inherently conservative nature of the majority voting strategy. 

\begin{minipage}{0.99\textwidth}
    \begin{minipage}[]{0.5999\textwidth}
         \begin{minipage}{0.3\linewidth}
            \includegraphics[width=\linewidth]{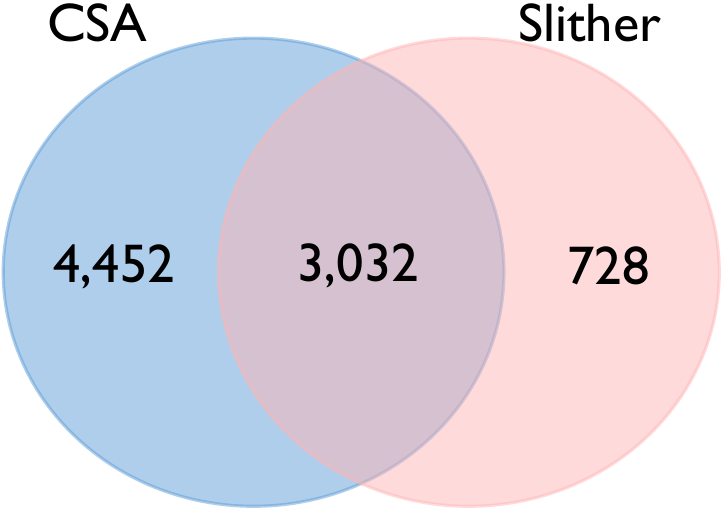}
            \\ \footnotesize (a) CSA \& Slither.
        \end{minipage}
        \begin{minipage}{0.3\linewidth}
            \includegraphics[width=\linewidth]{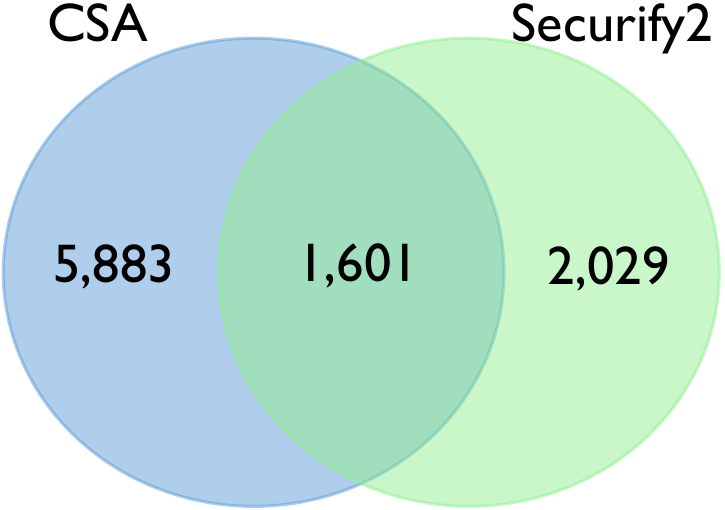}
            \\ \footnotesize (b) CSA \& Securify2.
        \end{minipage}
        \begin{minipage}{0.3\linewidth}
            \includegraphics[width=\linewidth]{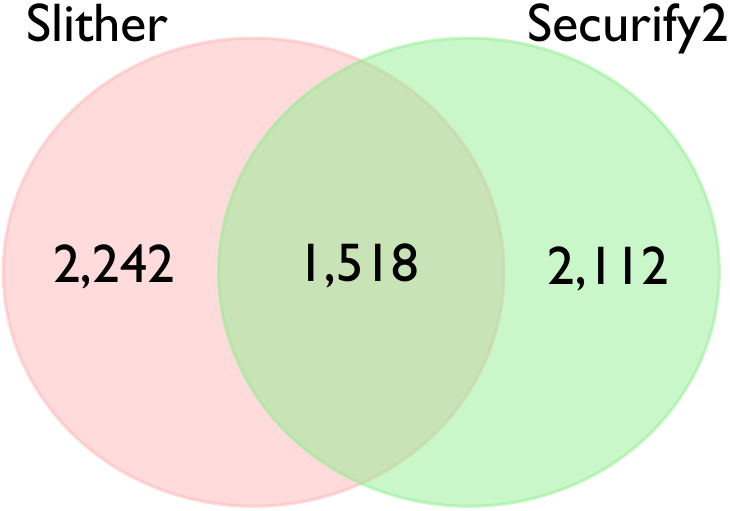}
            \\ \footnotesize (c) Slither \& Securify2.
        \end{minipage}
    \captionof{figure}{Combination of SAST tools.}
    \label{fig:combination}
  \end{minipage}
\hspace{5pt}
  \begin{minipage}[]{0.36\textwidth}
    \centering
    \captionof{table}{Analysis duration of each tool.}\label{tab: efficiency}
\resizebox{0.99\columnwidth}{!}{
\begin{tabular}{@{}ccr@{}}
\toprule
\multirow{2}{*}{\textbf{Tool}} & \multicolumn{2}{c}{\textbf{Duration}}     \\ \cmidrule(l){2-3}
                               & \multicolumn{1}{c}{\textbf{Average (s)}} & \multicolumn{1}{c}{\textbf{Total (s)}} \\ \midrule
\textbf{Securify2}             & 49.7                 & 446,741.0   \mybar{  5}          \\
\textbf{Slither}               & 16.6                 & 149,165.3   \mybar{  2}          \\
\textbf{SmartCheck}            & 112.7                & 1,012,242.0  \mybar{ 11}          \\
\textbf{CSA}                   & 17.5                 & 157,081.3   \mybar{  2}          \\
\textbf{Manticore}             & \textbf{1,547.6}      & 13,898,788.1 \mybar{100}          \\
\textbf{Mythril}               & 579.0                & 5,199,823.6  \mybar{ 35}          \\
\textbf{Osiris}                & 51.7                 & 464,279.6   \mybar{  5}          \\
\textbf{Oyente}                & 13.2                 & 118,301.1   \mybar{  1}          \\ \bottomrule
\end{tabular}
}
  \end{minipage}
\end{minipage}
\smallskip

\begin{tcolorbox}[size=title,opacityfill=0.1,boxsep=0mm]
    \textbf{Finding 7:} 
    The \textit{majority voting} among tools improves Precision to 36.10\%, increases the F1-score to 11.58\%, but decreases Recall to 6.90\%. 
    This approach reduces SAST tools' false alarms, emphasizing the precision-recall trade-off. 
\end{tcolorbox}





\subsection{RQ4: Efficiency Analysis}\label{rq4:efficiency}

As illustrated in Table~\ref{tab: efficiency}, it presents the average and total execution time of each tool when analyzing the selected smart contracts. On average, it takes about 298.5 seconds for each tool to analyze a contract. 
{Among SA tools,} Slither and CSA complete their analyses within 16.6 seconds and 17.5 seconds per contract on average, respectively, while SmartCheck tends to be less efficient, taking approximately 112.7 seconds on average. Notably, Securify2 demands more memory resources during analysis than other tools, a factor that users with resource constraints should consider.

{In contrast, SE tools require substantially more time to execute.} Manticore and Mythril have average analysis durations of $1,547.6s$ and $579.0s$, respectively. Oyente appears to be the fastest tool, taking an average of just $13.2s$ to analyze a contract. However, this is primarily due to the numerous compilation failures it encountered ($8,355$/$8,981$), as already shown in Table~\ref{tab:analysis_info}, which caused it to exit immediately when encountering such issues. Similarly, although Osiris is not as fast as Oyente, it is worth noting that it also had a significant number of compilation failures ($8,533$/$8,981$), which resulted in its relatively faster analysis time of $51.7s$.

In general, SA tools are faster than SE tools due to state traversal challenges faced by the latter. SA tools such as Securify2 and Slither, are faster due to their use of IR and the efficient program analysis techniques they employ. For instance, Securify2 transforms source code into an IR using the MLton compiler, an optimizer for the Standard ML programming language~\cite{MLton41:online}. Their faster execution time makes these tools more suitable for analyzing extensive contract sets or time-sensitive application scenarios. 
{SE tools, such as Mythril, often have longer execution times due to their technical nature, which explores all possible execution paths, leading to vast state space. This comprehensive exploration, while time-consuming, allows SE tools to uncover intricate vulnerabilities and edge cases, providing a deeper understanding of potential issues in smart contracts.}

\begin{tcolorbox}[size=title,opacityfill=0.1, boxsep=0mm]
            \noindent \textbf{Finding 8:}
            {SA tools generally outpace SE tools in terms of execution speed. Slither and CSA stand out with their swift scanning speed. In contrast, SE tools like Manticore and Mythril take considerably longer due to their exhaustive detection nature. While SA tools offer speed, SE tools provide depth, uncovering complex vulnerabilities by exploring all possible execution paths.}
\end{tcolorbox}

\section{Discussion}
\subsection{Implications}

\subsubsection{{For Tool Developers}}
Based on our findings, we suggest the following improvements for SAST tool developers:
\ding{172}\textit{ Handle compilation issues more robustly.} Our analysis in \S~\ref{sec:rq1-results} shows that these tools generally failed to perform analysis due to compilation issues. Future research should focus on developing tools that can accommodate a wider range of compiler versions and handle compilation issues more robustly. This may involve incorporating techniques such as partial compilation or on-the-fly compiler version switching. By enhancing the compatibility of these tools with various compiler versions, the effectiveness and reliability of smart contract analysis can be significantly improved. 
{\ding{173} \textit{Improve Precision by refining detection implementation.}}
As shown in \S~\ref{sec:rq2-results}, existing tools, including the commercial tool CSA, generally face a dilemma where they report numerous issues when analyzing smart contracts. 
{As depicted in \S~\ref{consistency}, one potential cause could be the heavy reliance on pattern matching for detection. The coverage and accuracy of these predefined rules can greatly influence effectiveness. 
Developers are expected to refine the detection rules by capturing the exact semantics of existing vulnerabilities rather than merely using intuitive pattern-matching. For instance, when detecting  \textit{Reentrancy} vulnerabilities, tools are expected to capture the state changes rather than a simple search for \texttt{call.value}. 
}
\ding{174} \textit{Focus on the interoperability of tools.} The findings in \S~\ref{consistency} demonstrate that some tools perform better in certain vulnerability categories than others. Therefore, it is essential to consider the interoperability of different tools, allowing users to combine the strengths of multiple tools for a more comprehensive security analysis. This can be achieved by adopting standardized input and output formats and developing frameworks that facilitate the integration of multiple tools. 
\ding{175} \textit{Improve the efficiency of tools.} Our study shows that some tools, such as Manticore analyzed contracts much slower than expected (1,547.6 seconds per contract), which can negatively impact their practical effectiveness. In particular, Manticore does not support a customized \texttt{max recursion depth} setting for users when compared to the other SE tools. To address this issue, developers should focus on optimizing tool performance by implementing more efficient algorithms, providing better resource management, and offering user-configurable settings for fine-tuning tool behavior based on specific requirements. 

\subsubsection{{For Researchers}}
We encourage future research in smart contract security to:
\ding{172} \textit{Propose a more comprehensive and up-to-date taxonomy and develop a diverse benchmark.} Taxonomies facilitate benchmark construction and objective evaluation in the context of smart contracts' development. As smart contract technologies evolve, new vulnerability types may emerge, necessitating a continuously updated, unified, and comprehensive taxonomy. Simultaneously, we encourage the development of a comprehensive and diverse benchmark for objectively evaluating SAST tools in line with the evolving taxonomy. 
{\ding{173} \textit{Call for a flexible IR design in SAST for smart contracts.} 
As discussed in \S~\ref{sec:rq2-results}, Slither and Securify2's compilation-based IR approach provides detailed analyses, while SmartCheck's direct transformation method, although resolving compilation challenges, might lack depth. IR profoundly affects the Precision and success rate of SAST tools, emphasizing the need for a balance between granularity and analysis success rate. To improve these tools, developers could explore hybrid IR strategies that combine the strengths of both compilation-based (IR) and direct transformation techniques. 
Continuous refinement based on feedback and benchmark performance, coupled with collaborative research, could pave the way for more robust IR methodologies that capture code intricacies without compromising the analysis success rate.} 
\ding{174} \textit{Better balance Recall and Precision in SAST tools for smart contracts.} Tools using SA generally achieve higher Recall but suffer from false positives, while SE tools exhibit better Precision but are limited by higher true positives due to their inherent technological limitations. Identifying effective strategies for balancing Recall and Precision presents a challenge for future research in the field of smart contract security analysis. 

\subsubsection{{For Practitioners}}
For practitioners working with smart contract SAST tools, we offer the following practical guidance:
\textit{\ding{172} Select tools carefully by considering their strengths and limitations.} Be aware that some tools may have limitations due to compilation issues and may not support the latest compiler versions. 
\ding{173} \textit{The strategic application of multiple SAST tools, based on their unique strengths, can cater to specific needs in vulnerability detection. }
As shown in \S~\ref{consistency}, tools demonstrate diverse strengths, with Securify2 excelling in detecting \textit{Denial of Service} vulnerabilities and Slither displaying proficiency in \textit{Reentrancy} issues.  
Importantly, the choice of combination strategy can greatly impact the Precision-Recall balance. 

\subsection{Threats to Validity}
\subsubsection{External Validity}~\label{external_threat}
\ding{172} The first threat concerns the generalizability of our evaluation results, specifically whether our findings can be applied to other SAST tools and datasets for Solidity smart contracts. We mitigated it by collecting existing benchmarks plus 113 real-world BNB projects, which contain validated vulnerabilities mapped and grouped into 45 unique vulnerability types in our taxonomy. Furthermore, we selected eight state-of-the-art tools (seven open-source and one commercial) that employ various techniques and are frequently used and/or evaluated in recent top-tier publications and prominent vendors. 
\ding{173} Another threat comes from the four collected benchmarks. As discussed in \S~\ref{benchmark}, we collected benchmarks with ``labeled vulnerability types'' and labeled vulnerable functions based on their claimed types. During this process, we identified and labeled unclaimed vulnerabilities, such as 1,611 new vulnerabilities in the SolidiFI Benchmark compared to their claimed 7,995. To address this, we thoroughly labeled additional vulnerabilities for a comprehensive evaluation of the tools' Precision. 
\ding{174} The last threat concerns the skewed distribution within our benchmark (\S~\ref{mapping}). 
This skew could potentially bias the overall effectiveness results. To mitigate it, we provide a detailed breakdown of the detection results on each vulnerability type in ~\Cref{tab:taxonomy}, aiming to offer a clearer insight into the tools' effectiveness.
\subsubsection{Internal Validity}~\label{internal_threat}
\ding{172} A possible threat arises from mistakes in labeling and mapping for ground truth and detection rules of the selected tools. To mitigate this, we collaborated with our industry partner and engaged three security experts to label the ground truth at the function level, a process that took a total of 7.5 person-months. Furthermore, three security auditing experts independently mapped the ground truth and detection rules, discussing any conflicts in mapping results until reaching an agreement, which required an additional 3.5 person-months. 
\ding{173} Another threat involves the runtime parameters used during the execution of the tools, particularly the one-hour timeout setting. 
We adopted this timeout based on recommendations in~\cite{ren_2021}. 
Notably, our in-depth analysis on Manticore in \S~\ref{sec:rq2-results} highlighted its practicality: despite facing numerous timeouts, a constraint-free rerun of Manticore on the SolidiFI benchmark led to analyzing only 36 out of 260 contracts, without any discoveries. 
This outcome, coupled with the analysis in \S~\ref{rq4:efficiency} and~\Cref{tab: efficiency}, affirms the one-hour limit's adequacy for most contracts.


\section{Conclusion and Future Work}
This paper presents a comprehensive evaluation of eight SAST tools for Solidity smart contracts including {coverage, effectiveness, consistency, and efficiency}. 
\revised{
We identify notable gaps in rule coverage, with tools struggling to comprehensively support the evolving range of Solidity vulnerabilities. Critical issues include high false positives in static analysis/source-code-based tools and symbolic execution/bytecode-based tools' limited path coverage. 
We highlight the urgent need for tool advancements to address compiler limitations and introduce more effective strategies, such as partial compilation or version adaptability. The varied effectiveness across vulnerability types suggests a potential for refining specialized tools. Emphasizing the need for improved detection methods, we suggest exploring hybrid analysis for complex vulnerabilities like reentrancy. 
Future research should consider improving the detection implementation by capturing the exact vulnerability patterns and exploring
sophisticated tool combination strategies, and efficiency enhancements (especially for Manticore) to bolster SAST tool performance. Our insights guide tool enhancement efforts, contributing to the security of the blockchain ecosystem. 
}



\begin{acks}
The authors would like to thank the anonymous reviewers for their constructive comments.  
This work is supported by the National Key R\&D Program of China under grant 2021ZD0114501, ECNU \& Huawei Trustworthiness Innovation Center, National Research Foundation, Singapore, the Cyber Security Agency under its National Cybersecurity R\&D Programme (NCRP25-P04-TAICeN), and the National Research Foundation Singapore and DSO National Laboratories under the AI Singapore Programme (AISG Award No: AISG2-RP-2020-019). Any opinions, findings, conclusions, or recommendations expressed in this material are those of the author(s) and do not reflect the views of the National Research Foundation, Singapore, and Cyber Security Agency of Singapore.
\end{acks}
\bibliographystyle{ACM-Reference-Format}
\bibliography{sample-base}

\end{document}